\documentclass[
 amsmath,amssymb,
 aps, 
prl,
twocolumn
]{revtex4-2}

\usepackage{graphicx}% Include figure files
\usepackage{dcolumn}% Align table columns on decimal point
\usepackage{bm}% bold math

\usepackage{siunitx}
\usepackage[version=4]{mhchem}
\usepackage{multirow}

\newcommand{\T}{^\mathsf{T}}

\newcommand{\atanTwo}{\mathrm{atan2}}

\renewcommand{\Re}{\operatorname{Re}}
\renewcommand{\Im}{\operatorname{Im}}

\begin{document}

\title{\textbf{Broadband Polarization Compensation with Link Segment Reconstruction for Quantum Optical Links} 
}% 

\author{Qingyu Shi}
 \email{Contact author: qy.shi@tum.de}
\author{Erwan Trad}%
\author{Julien Chénedé}
\author{Tobias Vogl}
 \email{Contact author: tobias.vogl@tum.de}
\affiliation{%
 Department of Computer Engineering, TUM School of Computation, Information and Technology, Technical University of Munich, 80333 Munich, Germany \\
 Munich Center for Quantum Science and Technology (MCQST), 80799 Munich, Germany
}%

\date{\today}

\begin{abstract}
Polarization-encoded quantum communication requires compensation of polarization transformations induced by the optical links. 
If the compensator is embedded between two channel segments, the transformations before and after the compensator must be treated separately. 
Moreover, standard three-wave-plate polarization controllers can become non-universal when their retardances deviate from their ideal values. 
To address these two challenges, we introduce a four-wave plate compensator that synthesizes arbitrary \(SO(3)\) polarization transformations over a broad wavelength range, and an eight-Stokes vector protocol that reconstructs the two link-segment Mueller matrices on either side of the compensator. 
Our experiment reveals that the four-plate sequence suppresses polarization-induced excess quantum bit error rate (QBER) to the sub-percent level at an operating wavelength more than \(100\,\mathrm{nm}\) from the design wavelength without further optimization. 
Combined with two auxiliary wavelengths, our scheme tracks the temperature-driven drift of a strongly wavelength-sensitive fiber spool while keeping the excess QBER below \(1\%\). 
These results support flexible compensator placement and wavelength channel selection, as well as non-interruptive polarization control in wavelength-division-multiplexed quantum optical links.
\end{abstract}

\maketitle

\textit{Introduction---}
Polarization is one of the most accessible photonic degrees of freedom for quantum communication because of its simple state preparation and analysis, compatibility with quantum key distribution (QKD) protocols such as BB84~\cite{Bennett14,Bennett92}, and has therefore been widely used in fiber and free-space demonstrations of quantum networks~\cite{Chen09,Wang14,Li18,Wengerowsky18,Agnesi20,Joshi20,Kupko20,Shi23,Luo24,Tan24,Zhou24,Ko25,Shi25,Zhang25,Talcott26}. 
In an optical link, however, polarization is generally not preserved.
Fiber birefringence, stress, bending, and component imperfections produce polarization transformations~\cite{Ulrich79}. 
They are, in general, wavelength dependent even in a static link and further drift under environmental perturbations~\cite{Chartier01,Luo24,Banner26}. 
For polarization-encoded quantum states, such uncompensated transformations lead directly to excess polarization error, reduced visibility, or loss of state fidelity at the receiver. 

A variety of polarization compensation and stabilization schemes have therefore been developed for quantum optical links. 
They are typically based on classical reference pulses or active feedback from measured quantum signals, and often formulated for a specific compensation geometry, with the compensator placed close to the detection stage~\cite{Chen09,Li18,Shi23}. 
In comparison, placing the compensator at an intermediate network node, rather than immediately before the detectors, can be useful when polarization control hardware is shared or co-located with other optical networking components. 
With this placement, however, the compensator cannot simply invert the end-to-end Mueller matrix, because the required transformation depends separately on the link segments before and after it. 
Although recent work has considered compensation geometries in which polarization transformations occur on both sides of the controller,  the role of such measurements as a tomography of the two unknown surrounding link transformations, and as a way to directly calculate the settings required for a middle-link compensator, has not been explicitly developed~\cite{Wu22,Tan24}.

At the same time, actual operating wavelengths may differ from the design wavelength of the polarization control elements. 
Retardance dispersion can reduce the accuracy of polarization compensation, especially for compensation schemes using wave plates, since the model assumes ideal quarter- and half-wave plates~\cite{Simon90,Bagini96,Wang07,Wang14}. 
This issue is particularly relevant for wavelength-division multiplexed (WDM) quantum network architectures, which are becoming increasingly important in scalable networks, where multiple users, frequency channels, or quantum--classical services share the same fiber infrastructure~\cite{Wengerowsky18,Joshi20,Shi25,Yadav25,Seo26,Talcott26}. 

In this Letter, motivated by these considerations, two core methods are developed: non-ideal four-wave plate synthesis of arbitrary polarization rotations over a broad wavelength range, and link-segment reconstruction for middle-link compensation. 
First, we address the synthesis problem. 
Although the standard Q-Q-H sequence is universal for ideal quarter- and half-wave plates~\cite{Simon90,Bagini96,Wang07}, we show that generic retardance errors prevent it from covering all \(SO(3)\) polarization rotations. 
We then introduce a Q-Q-Q-H compensator that can synthesize arbitrary \(SO(3)\) transformations under a conservative retardance error condition, allowing wavelength-flexible operation with non-ideal wave plates. 
Second, we introduce a link-segment reconstruction protocol for a compensator placed inside the optical link. 
The Mueller matrices before and after the compensator are determined from eight Stokes vector measurements, allowing the required middle-link compensation to be calculated directly. 
We combine the reconstruction and synthesis procedures, validate them at both near-design and off-design wavelengths, and finally demonstrate them in an auxiliary wavelength compensation experiment on a drifting fiber spool link. 
Together, these two methods provide a deterministic compensation framework for links in which the compensator is embedded between two unknown polarization transformations, and the operating wavelength does not necessarily coincide with the nominal wave plate design
wavelength. 

\textit{Synthesis---}
To implement the required compensation, we first address the synthesis of polarization transformations with non-ideal wave plates. 
We represent polarization states by normalized three-component Stokes vectors \(\Vec{S}\). 
A loss-normalized, non-depolarizing optical channel then acts on these vectors through a real \(3\times3\) matrix, \(\Vec{S}_{\rm out}=M\Vec{S}_{\rm in}\), with \(M\in SO(3)\), i.e., the
three-dimensional special orthogonal group~\cite{Aiello04,Goldstein17}. 
Synthesizing an arbitrary non-depolarizing polarization compensator is equivalent to synthesizing an arbitrary rotation on the Poincaré sphere. 
Note that polarization-dependent loss and depolarization are not treated as compensable errors in this Letter. 

It is known that three ideal wave plates, for example, in a Q-Q-H or Q-H-Q sequence, where Q and H denote quarter- and half-wave plates, respectively, are sufficient to synthesize an arbitrary $SO(3)$ polarization transformation~\cite{Simon90,Bagini96}. 
In the Mueller representation used here, this corresponds to constructing an arbitrary \(SO(3)\) matrix. 
This universality, however, assumes ideal retardances, namely \(\pi/2\) for each quarter-wave plate (QWP) and \(\pi\) for the half-wave plate (HWP). 
Fabrication tolerances and retardance dispersion make real wave plates deviate from these values, and the deviation generally increases when operated away from the design wavelength. 
Such non-ideal retardances make a generic Q-Q-H sequence no longer universal, as shown in Supplemental Material Section S1~\cite{SM}. 
The obstruction already appears for the simple class of rotations about the right-circular Stokes basis vector \(\Vec S_R=[0,0,1]\T\),
\[
M_R(\phi)=
\begin{bmatrix}
\cos\phi & \sin\phi & 0\\
-\sin\phi & \cos\phi & 0\\
0 & 0 & 1
\end{bmatrix},
\]
which leave \(\Vec S_R\) fixed and rotate the \(H\)-\(D\) plane. 
For generic retardance errors, only a finite set of such \(R\)-axis rotation angles can be realized, rather than arbitrary \(\phi\). 
Therefore \(R\)-axis rotations \(M_R(\phi)\) are not generally reachable.

To remove this reachability obstruction, we introduce an additional QWP before the Q-Q-H sequence, forming a Q-Q-Q-H compensator. 
The additional QWP acts as a preconditioning element. 
It changes the effective target transformation seen by the remaining Q-Q-H sequence and brings it into a reachable region. 
In the Supplemental Material Section S1~\cite{SM}, we show that this four-plate sequence can synthesize an arbitrary \(SO(3)\) Mueller matrix under the conservative sufficient condition
\begin{equation}
    |\Delta\delta_+| + |\Delta\delta_1| + |\Delta\delta_2| + |\Delta\delta_3|
    < \frac{\pi}{2},
    \label{eq:3-2-1}
\end{equation}
where \(\Delta\delta_+\) is the retardance error of the additional QWP, and \(\Delta\delta_1\), \(\Delta\delta_2\), and \(\Delta\delta_3\) are the retardance errors of the two QWPs and the HWP in the original Q-Q-H sequence, respectively.

The condition in Eq.~\eqref{eq:3-2-1} guarantees the existence of a solution, but does not by itself give the corresponding wave plate angles. 
The actual angles are computed with a Q-Q-Q-H angle-solving algorithm, described in the Supplemental Material Section S1~\cite{SM}. 
The algorithm analytically reduces the problem to a finite set of solution branches, and then performs a one-dimensional numerical minimization on each branch to select angles for which the generated Q-Q-Q-H Mueller matrix matches the target matrix.

For the commercial zero-order \(633\,\mathrm{nm}\) wave plates used in our experiment, using the retardance data from the manufacturer, we evaluate that the the sufficient condition is fulfilled over a broad wavelength interval, from \(535.4\,\mathrm{nm}\) to at least \(750\,\mathrm{nm}\)~\cite{SM}. 

\textit{Link-segment reconstruction---}
We next turn to the reconstruction problem for Mueller matrices of a compensator placed inside the optical link, rather than after all polarization distortion. 
We denote the source-to-compensator and compensator-to-detector transformations by \(M_\alpha\) and \(M_\beta\), respectively, and the wave plate transformation by \(M\), under the same \(SO(3)\) Mueller matrix model. The end-to-end transformation is then
\begin{equation}
    \Vec{S}_{\rm out}=M_\beta M M_\alpha \Vec{S}_{\rm in}.
\end{equation}

Perfect compensation requires $ M_\beta M M_\alpha = I $, and hence
\begin{equation}
    M = M_\beta^{\mathsf T} M_\alpha^{\mathsf T}.
    \label{eq:middle-link-compensation}
\end{equation}
Thus, when the compensator is placed in the middle of the optical link, the required compensating matrix is not the inverse of the uncompensated end-to-end transformation, i.e.\ , $(M_\beta M_\alpha)\T$, in general. 
Instead, the two link segments must be characterized separately.

Although \(M_\alpha\) and \(M_\beta\) cannot be accessed directly, they can be reconstructed from a small number of end-to-end measurements. 
We set the wave plate stack successively to four known transformations,
\begin{equation*}
    N_0=I,
\end{equation*}
and
\begin{equation*}
    N_1=
    \begin{bmatrix}
        1 & 0 & 0\\
        0 & 0 & -1\\
        0 & 1 & 0
    \end{bmatrix},
    N_2=
    \begin{bmatrix}
        0 & 0 & 1\\
        0 & 1 & 0\\
        -1 & 0 & 0
    \end{bmatrix},
    N_3=
    \begin{bmatrix}
        0 & -1 & 0\\
        1 & 0 & 0\\
        0 & 0 & 1
    \end{bmatrix}.
\end{equation*}
Here \(N_1\), \(N_2\), and \(N_3\) are rotations by \(\pi/2\) about the three Cartesian Stokes axes. 
In the experiment, each \(N_i\) was implemented with the Q-Q-Q-H compensator. 

For each setting \(N_i\), we measure the output Stokes vectors for two horizontal ($H$) and diagonal ($D$) input states,
\[
    \Vec{S}_H=[1,0,0]^{\mathsf T},
    \qquad
    \Vec{S}_D=[0,1,0]^{\mathsf T}.
\]
These two measurements determine the first two columns of the corresponding end-to-end Mueller matrix, while the third column is fixed by the \(SO(3)\) constraint. Denoting the two measured output vectors by \(\vec a_i\) and \(\vec b_i\), we write
\begin{equation}
    E_i
    =
    M_\beta N_i M_\alpha
    =
    \begin{bmatrix}
         \vec a_i & \vec b_i & \vec a_i\times \vec b_i
    \end{bmatrix}.
    \label{eq:end-to-end-matrix}
\end{equation}
In the experiment, the measured vector pair $(\vec a_i, \vec b_i)$ is orthonormalized before forming \(E_i\), see Supplemental Material Section S2~\cite{SM}. 

The reference setting \(N_0=I\) gives
\begin{equation}
    E_0=M_\beta M_\alpha .
\end{equation}
For \(i=1,2,3\), multiplying \(E_i\) by \(E_0^{\mathsf T}\) eliminates
\(M_\alpha\):
\begin{equation}
    C_i
    =
    E_i E_0^{\mathsf T}
    =
    M_\beta N_i M_\beta^{\mathsf T}.
    \label{eq:Ci-definition}
\end{equation}
Thus, \(C_i\) is a rotation by \(\pi/2\) about \(\vec u_i=M_\beta \hat e_i\), where \(\hat e_i\) is the corresponding Cartesian Stokes basis vector. 
We obtain \(\vec u_i\) from the antisymmetric part of \(C_i\):
\begin{equation}
    \vec u_i
    =
    \frac{1}{2}
    \begin{bmatrix}
        (C_i)_{zy}-(C_i)_{yz}\\
        (C_i)_{xz}-(C_i)_{zx}\\
        (C_i)_{yx}-(C_i)_{xy}
    \end{bmatrix},
    \qquad i=1,2,3.
    \label{eq:axis-extraction-main}
\end{equation}
The output-side link matrix is then reconstructed as
\begin{equation}
    M_\beta
    =
    \begin{bmatrix}
        \vec u_1 & \vec u_2 & \vec u_3
    \end{bmatrix}.
    \label{eq:Mbeta-reconstruction}
\end{equation}
In the ideal case, the reconstructed \(M_\beta\) is already orthonormal. 
Experimentally, small measurement and actuation errors can make the vectors $\vec u_i$ slightly nonorthogonal, so we project them onto \(SO(3)\) when estimating \(M_\beta\), see Supplemental Material Section S2~\cite{SM}. 

Finally, the input-side link matrix follows from
\begin{equation}
    M_\alpha
    =
    M_\beta^{\mathsf T} E_0 .
    \label{eq:Malpha-reconstruction}
\end{equation}
The compensating Mueller matrix to be implemented by the wave plate stack is then obtained from Eq.~\eqref{eq:middle-link-compensation}.

This protocol requires eight Stokes vector measurements in total: for each of the four wave plate settings \(N_i\) where \(i=0,1,2,3\), two input states \(H\) and \(D\) are evaluated. 
Within the non-depolarizing \(SO(3)\) model, these measurements provide the full information of polarization transformation on the two link segments and allow deterministic calculation of the required
compensation matrix.

\begin{figure}
    \centering
    \includegraphics[width=\linewidth]{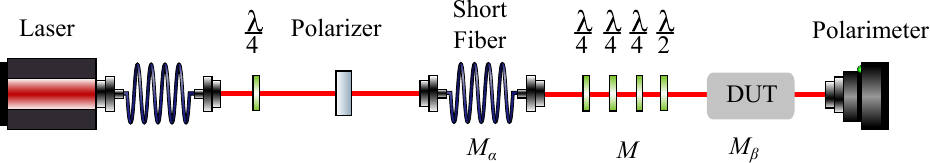}
    \caption{
    Experimental setup. 
    In the present implementation, the optical switch selects one fiber-coupled laser at a time.  
    The input state is prepared by a QWP followed by a linear polarizer on a motorized rotation mount. 
    The transformations before the compensator, through the compensator, and after the compensator are denoted by \(M_\alpha\), \(M\), and \(M_\beta\), respectively. 
    The compensator is a motorized stack of three QWPs followed by one HWP in the propagation direction. 
    DUT: device under test. 
    The output Stokes vectors are measured with a polarimeter.
    }
    \label{fig:exp_setup}
\end{figure}

Fig.~\ref{fig:exp_setup} shows the experimental implementation used for the two-step reconstruction and compensation procedure. 
The motorized Q-Q-Q-H compensator is placed between the input-side transformation \(M_\alpha\) and the output-side transformation \(M_\beta\). 
For each fixed wavelength, the compensation setting was obtained in two steps. 
First, we reconstructed \(M_\alpha\) and \(M_\beta\) using the eight-Stokes vector measurement protocol. 
Second, we calculated the Q-Q-Q-H wave plate angles required to implement the compensating matrix in Eq.~\eqref{eq:middle-link-compensation}. 

The compensated link was then benchmarked by scanning the azimuth \(\theta\) of a linear input polarization and measuring the output Stokes vector \(\Vec{S}'(\theta)\). 
For ideal compensation, the expected output is \(\Vec{S}(\theta)=[\cos2\theta,\sin2\theta,0]^{\mathsf T}\). 
We define the polarization-induced excess quantum bit error rate (QBER) as~\cite{Li18,Wu22}
\begin{equation}
    {\rm QBER}_{\rm pol}(\theta)
    =
    \frac{1-\Vec{S}(\theta)\cdot\Vec{S}'(\theta)}{2}.
    \label{eq:polarization-qber}
\end{equation}
This quantity is the polarization mismatch inferred from classical Stokes measurements in the form of QBER, not a full quantum-link QBER including detector noise, background photons, timing errors, or shot noise~\cite{Chen09,Xavier09,Agnesi20,Kupko20,Abasifard24,Luo24,Cholsuk25}. 
To characterize the compensation over different linear bases, we average \({\rm QBER}_{\rm pol}(\theta)\) over the scanned input azimuths. 

In the first experiment, we used zero-order wave plates designed for \(\SI{633}{nm}\) and tested the compensation at \(\SI{640}{nm}\) and \(\SI{515}{nm}\). 
Although the latter lies out of the wavelength interval over which Eq.~\eqref{eq:3-2-1} is fulfilled, the Q-Q-Q-H solver was checked numerically at this wavelength~\cite{SM}. 
This means that the reachability of compensation outside the interval guaranteed by the sufficient condition is not excluded. 
This experiment demonstrates the robustness of the wave plate construction when the operating wavelength differs from the design wavelength.
We also compared the Q-Q-Q-H sequence with a Q-Q-H baseline. 
For the Q-Q-H baseline, the compensation angles were calculated using the same measurement protocol but without the additional preconditioning QWP, and assuming the wave plates are ideal. 
The distortion element was a \(\SI{5}{m}\) single-mode fiber (SMF).

The results are summarized in Table~\ref{tab:table1}. 
\begin{table}[b]
\caption{\label{tab:table1}
Comparison of polarization-induced excess QBER after compensation with Q-Q-Q-H and Q-Q-H wave plate sequences. 
Input polarization azimuths were scanned from \(\SI{0}{\degree}\) to \(\SI{179}{\degree}\) in \(\SI{1}{\degree}\) steps on a \SI{5}{m} SMF. 
Values reported as \(x\pm y\) give the mean and standard deviation over the launched test polarization azimuths.
}
\begin{ruledtabular}
\begin{tabular}{ccc}
Wavelength (nm) & \multicolumn{2}{c}{Excess QBER (\%)}\\
\cline{2-3}
 & \shortstack{\\Q-Q-Q-H \\ (our method)} & \shortstack{Q-Q-H \\ (baseline)}\\
\colrule
640 & $0.051\pm0.050$ & $0.33\pm0.19$\\
515 & $0.41\pm0.23$  & $19.80\pm9.48$\\
\end{tabular}
\end{ruledtabular}
\end{table}
The Q-Q-Q-H sequence maintained sub-percent polarization-induced excess QBER at both wavelengths, whereas the Q-Q-H baseline degraded strongly at \(\SI{515}{nm}\). 
This further supports the analysis that the additional QWP mitigates the obstruction caused by a non-ideal three-plate sequence. 
The remaining residual error and its dependence on input azimuth can be attributed to finite extinction ratios, retardance calibration error, polarimeter measurement uncertainty, and wave plate positioning errors or drift. An independent characterization of the experimental baseline and relevant hardware limitations is provided in Supplemental Material Section S3~\cite{SM}.

In the second experiment, we applied the same Q-Q-Q-H reconstruction--synthesis procedure to the following four different link components to test its performance across different optical paths. 
This includes a \(1\,\mathrm{m}\) SMF segment, a \(5\,\mathrm{m}\) SMF mounted on a motorized fiber polarization controller, an optical switch module with SMF connections, and a \(100\,\mathrm{m}\) fiber spool. 
At \SI{635}{nm}, the resulting polarization-induced excess QBERs were \(0.055 \pm 0.043\%\), \(0.18 \pm 0.07\%\), \(0.13 \pm 0.09\%\), and \(0.033 \pm 0.027\%\), respectively.
Values reported as \(x\pm y\) give the mean and standard deviation over the launched test polarization azimuths.
These results show that the middle-link compensation procedure with four wave plates gives low residual polarization error for different link components.

\textit{Auxiliary-wavelength compensation---}
As an application, we finally apply the same framework and the present experimental configuration to a polarization compensation scheme with auxiliary wavelengths. 
In a WDM quantum link, auxiliary optical channels can co-propagate with the signal channel and monitor polarization changes without using the signal photons themselves. 
This can avoid some limitations of approaches that use the quantum signal directly or insert reference pulses in the time sequence~\cite{Chen09,Li18,Shi23,Tan24}, which may consume signal resources, depend on the communication protocol, or disturb the link during calibration. 

In wavelength-dependent links, auxiliary and signal wavelengths can experience different polarization transformations~\cite{Chartier01,Dong07,Tentori13,Steininger25,Banner26}. 
Polarization compensation with two classical side channels has already been demonstrated in polarization-encoded QKD~\cite{Xavier09}, where two side channels were used to address this wavelength dependence. 
Recent first-principles modeling of polarization-mode dispersion further emphasizes that the accuracy of side-channel compensation depends on the wavelength difference and the detailed fiber configuration~\cite{Banner26}.

We observed the same qualitative wavelength dependence in our laboratory tests. 
We represented the measured $SO(3)$ Mueller matrices using a locally continuous axis-angle representation, as described in Supplemental Material Section S4~\cite{SM}. 
Using the same setup as in Fig.~\ref{fig:exp_setup}, we repeated the eight-Stokes link-segment reconstruction protocol described above at multiple wavelengths to measure the Mueller matrix \(M_\beta\) of the devices under test (DUTs). 
The resulting spectral dependence is shown in Supplemental Material Section S5~\cite{SM}. 

For auxiliary compensation, two nearby auxiliary wavelengths were used to estimate the segment matrices at the signal wavelength, without replacing the quantum signal with a classical calibration probe. 
The two link segments were reconstructed at the auxiliary wavelengths and interpolated in the same axis-angle representation to estimate the Mueller matrices \(\hat M_{\alpha,s}\) and \(\hat M_{\beta,s}\) at the signal wavelength. 
This two-wavelength estimate could yield low polarization mismatch when the auxiliary wavelengths were sufficiently close to the signal.

The required compensating matrix implemented at the signal wavelength was then 
$
    \hat M =
    \hat M_{\beta,s}^{\mathsf T}
    \hat M_{\alpha,s}^{\mathsf T}.
$
We applied this method to a wavelength-sensitive test link. 
The device under test was a composite link consisting of a \(\SI{100}{m}\) SMF wound on a spool with a diameter of \SI{15}{cm} together with the \(\SI{5}{m}\) fiber-on-paddles DUT described above. 
The auxiliary lasers were centered at \(\SI{669}{nm}\) and \(\SI{671}{nm}\), with the signal wavelength at \(\SI{670}{nm}\). 
In this visible-wavelength testbed, these closely spaced red wavelengths were generated by tuning the current of a single diode laser, rather than by implementing a complete WDM system, due to the lack of mature implementations for visible light. 
For this spool configuration, the compensation gave polarization-induced excess QBER values between \(\SI{0.15}{\percent}\) and \(\SI{0.5}{\percent}\), depending on the fiber state and paddle settings. 

Using this configuration, we then tested the stability of the auxiliary compensation under temperature variation. 
The \(\SI{100}{m}\) fiber spool was placed in a home-built temperature-controlled enclosure, and the temperature was varied between \(\SI{13}{\degreeCelsius}\) and \(\SI{30}{\degreeCelsius}\) with a period of $4$ hours. 
This temperature range was chosen based on the expected temperature range for a buried fiber in Munich, and the limitations of the enclosure, see Supplemental Material Section S6~\cite{SM,DWD26}. 

This measurement was designed as a step towards non-interruptive auxiliary wavelength tracking of polarization drift, without returning to a full recalibration. 
A full repetition of the eight-Stokes reconstruction would require applying several test transformations with the compensator and would therefore temporarily disturb the compensated link.
Instead, we kept the compensator near its operating point and used the two auxiliary wavelengths to guide local wave plate updates using a scalar feedback metric derived from the measured Stokes vectors, see Supplemental Material Section S7~\cite{SM}.

\begin{figure}
 \centering
 \includegraphics[width=\linewidth]{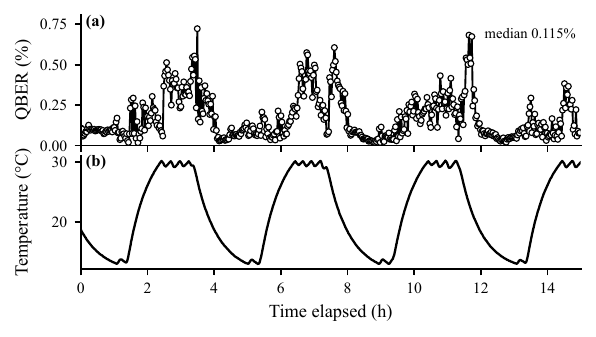}
 \caption{
 Long-term monitoring of a \(\SI{100}{m}\) fiber spool with active auxiliary wavelength polarization compensation. 
 The auxiliary wavelengths were \(\SI{669}{nm}\) and \(\SI{671}{nm}\), and the signal wavelength was \(\SI{670}{nm}\). 
 ~\textbf{(a)} Polarization-induced excess QBER during the measurement interval. 
 Each symbol represents one feedback/optimization iteration. 
 ~\textbf{(b)} Temperature recorded by the home-built temperature-controlled enclosure during the same interval. 
 }
 \label{fig:fiber_monitor_qber_temperature}
\end{figure}

As shown in Fig.~\ref{fig:fiber_monitor_qber_temperature}, the auxiliary wavelength feedback maintained low residual polarization mismatch throughout the measurement. 
The median excess QBER was \(\SI{0.115}{\percent}\), and the excess QBER remained below \(\SI{1}{\percent}\) over the recorded interval. 
The residual peaks can be attributed mainly to tracking latency in our implementation, especially during rapid temperature changes and near the high-temperature part of the cycle, where the link's polarization transformation is more temperature-sensitive \cite{SM}. 

The auxiliary--signal wavelength spacing required for this interpolation was device-dependent. 
For the less wavelength-sensitive fiber-on-paddles configuration, with the motorized polarization controller at its home position, auxiliary wavelengths at \(\SI{658.3}{nm}\) and \(\SI{686.5}{nm}\) gave an excess QBER of \(0.33\pm0.22\%\) at the signal wavelength \(\SI{672.3}{nm}\). 
In contrast, the same widely separated auxiliary wavelengths were not reliable for the highly dispersive composite paddle--spool DUT mentioned above. 
The residual excess QBER ranged from about \(0.7\%\) to \(15\%\), depending on the fiber state and paddle settings. 
This is consistent with the stronger spectral variation expected from a tightly wound fiber spool, where bend-induced retardance and its wavelength sensitivity increase with bend curvature~\cite{Ulrich79}. 

The close-wavelength experiment can be interpreted as a scale test of the auxiliary signal spacing required for reliable interpolation in a wavelength-sensitive link. 
The \(\SI{2}{nm}\) separation used around \(\SI{670}{nm}\) corresponds to a relative spacing of \(3.0\times10^{-3}\). This would correspond to \(\SI{4.6}{nm}\) when migrating to a working wavelength of \SI{1550}{nm}. 
Since a \(\SI{100}{GHz}\) dense-WDM channel spacing near the telecom C band is approximately \(\SI{0.8}{nm}\) \cite{ITU20}, the demonstrated relative spacing is on the order of six standard channel intervals. 
Links with weaker spectral variation could tolerate more widely separated auxiliary probes, leaving more room for intermediate WDM channels.

\textit{Conclusion---}
We have demonstrated a deterministic polarization compensation framework for polarization-sensitive optical links in a geometry in which the compensator is embedded inside the channel. 
The method combines link-segment Mueller matrix reconstruction with non-ideal wave plate synthesis, allowing the input- and output-side transformations to be tomographically estimated and the middle-link inverse to be calculated. 
The four-plate Q-Q-Q-H compensator tolerates retardance dispersion and gives significantly lower polarization-induced excess QBER than a Q-Q-H baseline at off-design wavelengths. 
The same reconstruction--synthesis procedure also gives low residual polarization error for different link components, including fiber segments with different lengths and bend conditions, as well as an optical switch. 
These results support flexible compensator placement, such as at an intermediate network node or within a reconfigurable switching module, as well as wavelength-flexible operation with non-ideal wave plates in quantum optical links.

We also used closely spaced auxiliary wavelengths to track temperature-driven polarization drift without repeating the full segment reconstruction, maintaining the polarization-induced excess QBER below \(1\%\) on a strongly wavelength-dependent \(100\,\mathrm{m}\) fiber spool link. 
The small bend radius of the spool and the relatively rapid temperature changes make this a severe test case, showing that the auxiliary spacing must be chosen according to the spectral variation of the deployed link. 
These results motivate telecom-band implementations of WDM-based auxiliary wavelength polarization compensation schemes with integrated multiplexing, faster polarimetry, and faster actuation.

\textit{Acknowledgements---}
This research is part of the Munich Quantum Valley, which is supported by the Bavarian state government with funds from the Hightech Agenda Bayern Plus. 
This work was funded by the Deutsche Forschungsgemeinschaft (DFG, German Research Foundation) under Germany's Excellence Strategy- EXC-2111-390814868 (MCQST). 
The authors acknowledge support from the Federal Ministry of Research, Technology and Space (BMFTR) under grant number 13N16292 (ATOMIQS). 

\textit{Data availability---}
The data are available from the authors upon reasonable request.

\textit{Notes---}
We acknowledge the use of Microsoft Copilot with the GPT-5 reasoning model to assist in improving Python scripts used for figure generation and in revising selected passages of the manuscript for readability. All scientific interpretations, data analyses, and final writing decisions are made by the authors.

\nocite{*}

\bibliography{paper01}

\clearpage
\onecolumngrid

\setcounter{secnumdepth}{3}

\setcounter{section}{0}
\setcounter{subsection}{0}
\setcounter{subsubsection}{0}
\setcounter{equation}{0}
\setcounter{figure}{0}
\setcounter{table}{0}

\renewcommand{\thesection}{S\arabic{section}}
\renewcommand{\theequation}{S\arabic{equation}}
\renewcommand{\thefigure}{S\arabic{figure}}
\renewcommand{\thetable}{S\arabic{table}}

% Supplemental Material title
\begin{center}
{\large\bfseries Supplemental Material}
\end{center}

\vspace{1em}

\section{Constructing an arbitrary $SO(3)$ Mueller matrix}

As discussed in the main text, we aim to use a wave plate stack to implement a Mueller matrix that compensates polarization transformations in an optical link. 
Throughout this work, we consider polarization transformations that act as rotations on the Poincaré sphere. 
We denote the normalized three-component Stokes vector by $\Vec{S}$. 
Polarization independent attenuation only rescales the total intensity and is removed by normalization. 
Polarization dependent loss and depolarization are not included in the present model. 
Under these assumptions, a Mueller matrix is represented by a real special orthogonal matrix $M\in SO(3)$, such that \cite{Aiello04,Goldstein17}
\begin{equation}
    \Vec{S}_{\mathrm{out}}=M\Vec{S}_{\mathrm{in}}.
\end{equation}

Constructing an arbitrary \(SO(3)\) Mueller matrix is equivalent to constructing a compensating transformation that maps distorted Stokes vectors back to their original positions on the Poincaré sphere. 
The basis vectors
\begin{equation}
\Vec{S}_H=(1,0,0)\T,\qquad 
\Vec{S}_D=(0,1,0)\T,\qquad 
\Vec{S}_R=(0,0,1)\T,
\end{equation}
correspond to horizontal (\(H\)), diagonal (\(D\)), and right-circular (\(R\)) polarization, respectively.

Suppose that for input polarizations \(H\), \(D\), and \(R\), the measured Stokes vectors at a detector after the distortion are $\Vec{S}_{H0}, \Vec{S}_{D0}, \Vec{S}_{R0}.$
The Mueller matrix describing this distortion is then
\begin{equation}
M_\alpha=
\begin{bmatrix}
\Vec{S}_{H0} & \Vec{S}_{D0} & \Vec{S}_{R0}
\end{bmatrix},
\end{equation}
because the three input Stokes vectors form the Cartesian basis. 
If the compensation element is placed after the distortion and immediately before the detection device, its Mueller matrix \(M\) should satisfy $M M_\alpha=I$.
Since \(M_\alpha\in SO(3)\), its inverse is its transpose, and therefore
\begin{equation}
M=M_\alpha^{-1}=M_\alpha^{\mathsf T}
=
\begin{bmatrix}
\Vec{S}_{H0} & \Vec{S}_{D0} & \Vec{S}_{R0}
\end{bmatrix}^{\mathsf T}.
\end{equation}
Conversely, constructing a desired \(SO(3)\) Mueller matrix \(M\) can be seen as constructing the compensating transformation for a distortion whose Mueller matrix is \(M^{\mathsf T}\). 

In this section, we first consider a standard Q-Q-H wave plate sequence, consisting of two quarter-wave plates (QWPs) followed by one half-wave plate (HWP), and show that retardance errors make this three-plate sequence non-universal. 
We then show that under a given conservative retardance error bound, adding one extra QWP restores arbitrary \(SO(3)\) reachability, and demonstrate the corresponding angle solving algorithm at the same time. 
Finally, we evaluate the acceptable wavelength range for the commercial wave plates used in the experiment required by this conservative bound.

\subsection{Non-ideal Q-Q-H combination}

We first consider a wave plate compensator with a Q-Q-H sequence \cite{Simon90,Bagini96}. The two QWPs have nominal retardance \(\pi/2\), and the HWP has nominal retardance \(\pi\). 
At the operating wavelength, however, the actual retardances may deviate from these ideal values. 
We determine which \(SO(3)\) Mueller matrices can be realized by such a non-ideal Q-Q-H sequence, and equivalently, which polarization distortions can be compensated.

The optical elements are numbered according to the propagation direction of light. 
We denote the slow-axis angles of the first, second, and third wave plates by \(\theta_1\), \(\theta_2\), and \(\theta_3\), respectively. 
The first two plates are nominal QWPs, and the third plate is a nominal HWP, with actual retardances
\begin{equation}
\delta_1=\frac{\pi}{2}+\Delta\delta_1,\qquad
\delta_2=\frac{\pi}{2}+\Delta\delta_2,\qquad
\delta_3=\pi+\Delta\delta_3.
\end{equation}
The corresponding Mueller matrices are denoted by \(M_1\), \(M_2\), and \(M_3\). 
Thus, the total Mueller matrix of the Q-Q-H sequence is
\begin{equation}
M_{\mathrm{QQH}}=M_3M_2M_1.
\end{equation}

To characterize the reachable set of the Q-Q-H sequence, we assume that compensation has succeeded, and propagate the target basis states $H, D, R$ backward through the compensator, one wave plate at a time. 

At the detector, successful compensation requires the three basis states to be restored to
\begin{equation}
\Vec{S}_{H3}=[1,0,0]^{\mathsf T},\qquad
\Vec{S}_{D3}=[0,1,0]^{\mathsf T},\qquad
\Vec{S}_{R3}=[0,0,1]^{\mathsf T}.
\end{equation}
We first propagate these states one step backward through the third wave plate, which is the HWP.
With the convention used here, the inverse Mueller matrix of the non-ideal HWP is
\begin{equation}
    M_3^{-1} = \begin{bmatrix}
        \cos^2 2\theta_3 - \sin^2 2\theta_3 \cos \Delta\delta_3 & \cos 2\theta_3 \sin 2\theta_3 (1 + \cos \Delta \delta_3) & \sin 2\theta_3 \sin \Delta\delta_3 \\
        \cos 2\theta_3 \sin 2\theta_3 (1+\cos\Delta\delta_3) & -\cos^2 2\theta_3 \cos \Delta\delta_3 + \sin^2 2\theta_3 & -\cos 2\theta_3 \sin \Delta \delta_3 \\
        -\sin 2\theta_3 \sin \Delta \delta_3 & \cos 2\theta_3 \sin \Delta\delta_3 & -\cos\Delta\delta_3
    \end{bmatrix}. 
    \label{eq:supp-HWP_inverse}
\end{equation}

Since
\begin{equation}
M_3
\begin{bmatrix}
\Vec{S}_{H2} & \Vec{S}_{D2} & \Vec{S}_{R2}
\end{bmatrix}
=
\begin{bmatrix}
\Vec{S}_{H3} & \Vec{S}_{D3} & \Vec{S}_{R3}
\end{bmatrix}
=
I_3,
\end{equation}
the Stokes vectors immediately before the HWP are
\begin{equation}
\begin{bmatrix}
\Vec{S}_{H2} & \Vec{S}_{D2} & \Vec{S}_{R2}
\end{bmatrix}
=
M_3^{-1}.
\end{equation}
Thus, \(\Vec{S}_{H2}\), \(\Vec{S}_{D2}\), and \(\Vec{S}_{R2}\) are the three columns of \(M_3^{-1}\).

For an ideal HWP, \(\Delta\delta_3=0\), Eq.~\eqref{eq:supp-HWP_inverse} reduces to a rotation by \(\pi\) about the axis $(\cos 2\theta_3,\sin 2\theta_3,0)^{\mathsf T}$.
In this ideal case, the \(H\)- and \(D\)-components remain on a circle with unchanged radius, while the \(R\)-component changes sign.

We express the normalized Stokes vector in angular coordinates on the Poincaré sphere, 
\begin{equation}
\Vec{S}
=
[ \cos 2\psi\cos 2\chi,~\sin 2\psi\cos 2\chi,~\sin 2\chi]\T,
\end{equation}
where \(\psi\) is the polarization azimuth and \(\chi\) is the ellipticity angle. 
Unless otherwise stated, they are not restricted to a principal branch. 
Two angular pairs are regarded as equivalent if they give the same Cartesian vector \(\Vec{S}\). 
The coordinate \(2\psi\) is defined modulo \(2\pi\), and when it is computed from Cartesian components, we use the convention
\begin{equation}
2\psi=\operatorname{atan2} (S_y,S_x).
\end{equation}

The vector \(\Vec{S}_{R2}\), corresponding to the third column of \(M_3^{-1}\), can then be represented by
\begin{align}
2\chi_{R2} &= -\frac{\pi}{2}+\Delta\delta_3, \nonumber\\
2\psi_{R2} &= 2\theta_3-\frac{\pi}{2}.
\label{eq:supp-choice_R2}
\end{align}

In the ideal limit \(\Delta\delta_3=0\), this vector lies at the pole \((0,0,-1)^{\mathsf T}\), where the azimuth is not unique, but the choice in Eq.~\eqref{eq:supp-choice_R2} is still a valid parametrization.

Because \(M_3^{-1}\in SO(3)\), its columns are mutually orthonormal. 
Therefore, \(\Vec{S}_{H2}\) and \(\Vec{S}_{D2}\) lie in the plane through the origin whose normal vector is \(\Vec{S}_{R2}\). 
We define the equatorial HWP axis
\begin{equation}
\Vec{A} = [\cos2\theta_3,~\sin2\theta_3,~0]\T,
\end{equation}
which also lies in this plane since \(\Vec{A}\cdot\Vec{S}_{R2}=0\).

Let \(\eta\) denote the signed angle in this normal plane measured from \(\Vec{A}\). 
From the first two columns of Eq.~\eqref{eq:supp-HWP_inverse}, the angles corresponding to \(\Vec{S}_{H2}\) and \(\Vec{S}_{D2}\) are
\begin{equation}
\eta_{H2}=2\theta_3,\qquad
\eta_{D2}=2\theta_3-\frac{\pi}{2}.
\end{equation}
When these vectors are projected onto the \(H\)-\(D\) plane, their components along \(\Vec{A}\) remain unchanged, whereas the perpendicular components are multiplied by \(\cos\Delta\delta_3\). 
Hence, the projected angle relative to \(\Vec{A}\) is
\begin{equation}
\eta_{\mathrm{proj}}
=
\atanTwo\left(\sin\eta\,\cos\Delta\delta_3,\,\cos\eta
\right).
\end{equation}
Therefore,
    \begin{align}
        2\chi_{H2} & = \arcsin(-\sin2\theta_3 \sin\Delta\delta_3), \nonumber \\
        2\psi_{H2} & = \atanTwo(\sin2\theta_3 \cos\Delta\delta_3, \cos 2\theta_3) + 2\theta_3, \nonumber \\
        2\chi_{D2} & = \arcsin(\cos2\theta_3 \sin\Delta\delta_3), \nonumber \\
        2\psi_{D2} & = \atanTwo(-\cos2\theta_3 \cos\Delta\delta_3, \sin 2\theta_3) + 2\theta_3. 
    \end{align}

We next propagate the vectors backward through the second plate, which is a nominal QWP. 
Its inverse Mueller matrix is
\begin{equation}
    M_2^{-1} =
    \begin{bmatrix}
        \cos^2 2\theta_2 - \sin^2 2\theta_2 \sin \Delta\delta_2
        &
        \cos 2\theta_2 \sin 2\theta_2 (1 + \sin \Delta \delta_2)
        &
        -\sin 2\theta_2 \cos \Delta\delta_2
        \\
        \cos 2\theta_2 \sin 2\theta_2 (1+\sin\Delta\delta_2)
        &
        -\cos^2 2\theta_2 \sin \Delta\delta_2 + \sin^2 2\theta_2
        &
        \cos 2\theta_2 \cos \Delta \delta_2
        \\
        \sin 2\theta_2 \cos \Delta \delta_2
        &
        -\cos 2\theta_2 \cos \Delta\delta_2
        &
        -\sin\Delta\delta_2
    \end{bmatrix}.
\end{equation}
The vector corresponding to the \(R\) basis state before the second QWP is
\begin{equation}
\Vec{S}_{R1}=M_2^{-1}\Vec{S}_{R2}.
\end{equation}
Defining
\begin{equation}
D=2\theta_3-2\theta_2,
\end{equation}
we obtain
\begin{equation}
    \Vec{S}_{R1} =
    \begin{bmatrix}
        \sin \Delta\delta_3 \left[ \cos 2\theta_2 \sin D - \sin\Delta\delta_2 \sin 2\theta_2 \cos D \right] + \cos \Delta\delta_3 \sin 2\theta_2 \cos \Delta\delta_2 \\
        \sin \Delta\delta_3 \left[ \sin 2\theta_2 \sin D + \sin\Delta\delta_2  \cos 2\theta_2 \cos D \right] - \cos\Delta\delta_3 \cos 2\theta_2 \cos \Delta\delta_2 \\
        \cos D \sin\Delta\delta_3 \cos\Delta\delta_2 + \cos\Delta\delta_3 \sin\Delta\delta_2
    \end{bmatrix}.
\end{equation}

The parameters \((\theta_2, D)\) can be regarded as the independent variables at this step, since \(\theta_3=\theta_2+D/2\). 
The \(x\) and \(y\) components can be written as a rotation by \(2\theta_2\),
\begin{equation}
\begin{bmatrix}
S_{R1,x}\\
S_{R1,y}
\end{bmatrix}
=
\begin{bmatrix}
\cos2\theta_2 & -\sin2\theta_2\\
\sin2\theta_2 & \cos2\theta_2
\end{bmatrix}
\begin{bmatrix}
X_{R1}\\
Y_{R1}
\end{bmatrix},
\end{equation}
where
\begin{align}
X_{R1}
&=
\sin\Delta\delta_3 \sin D, \nonumber \\
Y_{R1}
&=
\cos D\,\sin\Delta\delta_3\sin\Delta\delta_2
-
\cos\Delta\delta_3\cos\Delta\delta_2.
\end{align}
The \(z\) component is
\begin{equation}
S_{R1,z} = Z_{R1}
=
\cos D\,\sin\Delta\delta_3\cos\Delta\delta_2
+
\cos\Delta\delta_3\sin\Delta\delta_2.
\end{equation}
Therefore,
\begin{align}
        2\chi_{R1}
        &=
        \arcsin Z_{R1}, \nonumber \\
        2\psi_{R1}
        &=
        \operatorname{atan2} (Y_{R1},X_{R1})+2\theta_2. 
\end{align}
Equivalently,
\begin{align}
        2\chi_{R1}
        &=
        \arcsin\left(\cos D \sin\Delta\delta_3 \cos\Delta\delta_2 + \cos\Delta\delta_3 \sin\Delta\delta_2\right), \label{eq:supp-chi-R1}  \\
        2\psi_{R1}
        &=
        \operatorname{atan2} \left(
        \cos D \sin\Delta\delta_3 \sin\Delta\delta_2
        -
        \cos \Delta\delta_3 \cos\Delta\delta_2,
        \,
        \sin\Delta\delta_3 \sin D
        \right)
        +2\theta_2 .\label{eq:supp-psi-R1}
\end{align}

The first plate is also a nominal QWP, so \(M_1^{-1}\) is obtained from \(M_2^{-1}\) by replacing \((\theta_2,\Delta\delta_2)\) with \((\theta_1,\Delta\delta_1)\). 
Applying \(M_1^{-1}\) to \(\Vec{S}_{R1}\), and defining
\begin{equation}
B=2\theta_1-2\psi_{R1},
\end{equation}
gives
\begin{equation}
    \Vec{S}_{R0} = \begin{bmatrix}
        \cos2\chi_{R1} \left[ \cos 2\theta_1 \cos B - \sin \Delta\delta_1 \sin 2\theta_1 \sin B \right] - \sin 2\theta_1 \cos \Delta\delta_1 \sin 2\chi_{R1} \\
        \cos 2\chi_{R1} \left[ \sin 2\theta_1 \cos B + \sin\Delta\delta_1 \cos 2\theta_1 \sin B \right] + \cos 2\theta_1 \cos \Delta\delta_1 \sin 2\chi_{R1} \\
        \cos \Delta\delta_1 \cos 2\chi_{R1} \sin B - \sin \Delta\delta_1 \sin 2\chi_{R1}
    \end{bmatrix}.
    \label{eq:supp-SR0-vector}
\end{equation}
The corresponding azimuth is
\begin{equation}
    2\psi_{R0}
    =
    \operatorname{atan2}
    \left(
    \cos 2\chi_{R1} \sin\Delta\delta_1 \sin B
    +
    \cos\Delta\delta_1 \sin 2\chi_{R1},
    \,
    \cos 2\chi_{R1} \cos B
    \right)
    +2\theta_1 .
    \label{eq:supp-psi-R0}
\end{equation}

We need to find wave plate angles \(\theta_1,\theta_2,\theta_3\) such that the total Mueller matrix of the Q-Q-H combination maps the distorted $H,D,R$ Cartesian basis vectors back.~
Equivalently, the total Q-Q-H Mueller matrix
$M_{\mathrm{QQH}}=M_3M_2M_1$
should satisfy
\begin{equation}
M_{\mathrm{QQH}}=M,
\qquad
M=[\Vec{S}_{H0},\Vec{S}_{D0},\Vec{S}_{R0}]^{\mathsf T}.
\end{equation}
We first use the \(R\)-basis vector to generate candidate angles, and then verify which of them can construct the target matrix. 

The vector \(\Vec{S}_{R0}\) should be mapped to $\Vec{S}_{R3}=[0,0,1]^{\mathsf T}.$
From the \(z\)-component of Eq.~\eqref{eq:supp-SR0-vector}, we obtain
\begin{equation}
    \sin B =
    \frac{S_{R0,z}+\sin\Delta\delta_1 \sin 2\chi_{R1}}
    {\cos\Delta\delta_1 \cos 2\chi_{R1}},
    \label{eq:supp-sinB}
\end{equation}
provided that \(\cos\Delta\delta_1\cos2\chi_{R1}\neq0\). 
If the denominator in Eq.~\eqref{eq:supp-sinB} vanishes, the corresponding singular case is treated separately. 

For a given value of \(D\), Eq.~\eqref{eq:supp-sinB} gives zero, one, or two solutions for \(B\) in a \(2\pi\) period, depending on whether the absolute value of the right-hand side is larger than, equal to, or smaller than unity. 
If a value of \(D\) yields a solution, at least one valid branch of \(B\) must exist. 

For each admissible branch, the condition to recover \(R\)-axis is solved analytically, leaving only the relative angle $D=2\theta_3-2\theta_2$ to be searched numerically in order to match the target Mueller matrix. 
The procedure is as follows.  

\begin{enumerate}
    \item Choose a trial value of \(D\in[-\pi,\pi)\).

    \item Calculate \(2\chi_{R1}\) from Eq.~\eqref{eq:supp-chi-R1}:
    \begin{equation}
    2\chi_{R1}
    =
    \arcsin\left(
    \cos D \sin\Delta\delta_3 \cos\Delta\delta_2
    +
    \cos\Delta\delta_3 \sin\Delta\delta_2
    \right).
    \end{equation}

    \item Solve Eq.~\eqref{eq:supp-sinB} for all admissible branches of \(B\). For each branch \(b\), continue the following steps separately.

    \item For the selected branch \(B_b(D)\), calculate \(\theta_1\) from Eq.~\eqref{eq:supp-psi-R0}:
    \begin{equation}
        \theta_1^{(b)}(D)
        =
        \psi_{R0}
        -
        \frac{1}{2}
        \operatorname{atan2}
        \left(
        \cos 2\chi_{R1}\sin\Delta\delta_1\sin B_b
        +
        \cos\Delta\delta_1\sin 2\chi_{R1},
        \,
        \cos 2\chi_{R1}\cos B_b
        \right),
        \label{eq:supp-theta1-solver}
    \end{equation}
    where $2\psi_{R0} = \operatorname{atan2} (S_{R0,y},S_{R0,x}).$ 

    \item Apply the first QWP to obtain
    \begin{equation}
    \Vec{S}_{R1}
    =
    M_1\Vec{S}_{R0},
    \end{equation}
    and calculate its azimuthal coordinate
    \begin{equation}
    2\psi_{R1}
    =
    \operatorname{atan2} (S_{R1,y},S_{R1,x}).
    \end{equation}

    \item Calculate \(\theta_2\) from Eq.~\eqref{eq:supp-psi-R1}:
    \begin{equation}
        \theta_2^{(b)}(D)
        =
        \psi_{R1}
        -
        \frac{1}{2}
        \operatorname{atan2}
        \left(
        \sin\Delta\delta_3\sin\Delta\delta_2\cos D
        -
        \cos \Delta\delta_3\cos\Delta\delta_2,
        \,
        \sin\Delta\delta_3\sin D
        \right).
        \label{eq:supp-theta2-solver}
    \end{equation}

    \item Calculate
    \begin{equation}
    \theta_3^{(b)}(D)
    =
    \theta_2^{(b)}(D)
    +
    \frac{D}{2}.
    \end{equation}

    \item Build the Mueller matrix of the Q-Q-H sequence $ M_{\mathrm{QQH}} = M_3M_2M_1$ using the candidate angles $\theta_1^{(b)}(D), \theta_2^{(b)}(D), \theta_3^{(b)}(D)$.

    \item Evaluate the residual
    \begin{equation}
    F_b(D)
    =
    \left\|
    M_{\mathrm{QQH}}
    -
    M
    \right\|_{\mathrm F}.
    \end{equation}

    \item For each branch \(b\), numerically solve the scalar equation $F_b(D)=0$ for \(D\in[-\pi,\pi)\). 
    A valid solution on branch \(b\) exists only when this equation has a root. 
    We locate the root by minimizing the residual \(F_b(D)\) and accept the solution only when the residual is below a specified tolerance. 
    Each branch is treated separately and may have multiple solutions.
\end{enumerate}

\subsection{Necessity of the additional QWP in the Q-Q-Q-H sequence}

We now show why the three-plate Q-Q-H sequence is generally not sufficient to synthesize arbitrary \(SO(3)\) transformations when the retardances are non-ideal. 
To demonstrate this, we look for simple examples where the Q-Q-H angle-solving procedure has no solution.

We first examine the existence condition for the branch variable \(B\) in Eq.~\eqref{eq:supp-sinB}. 
In the non-singular case \(\cos\Delta\delta_1\cos2\chi_{R1}\neq0\), a real solution for \(B\) exists only if
\begin{equation}
    \left|
    \frac{S_{R0,z}+\sin\Delta\delta_1\sin2\chi_{R1}}
    {\cos\Delta\delta_1\cos2\chi_{R1}}
    \right|
    \leq 1 .
\end{equation}
Squaring both sides gives
\begin{equation}
    \left(S_{R0,z}+\sin\Delta\delta_1\sin2\chi_{R1}\right)^2
    \leq
    \cos^2\Delta\delta_1\cos^2 2\chi_{R1}.
\end{equation}
Using \(\cos^2 2\chi_{R1}=1-\sin^2 2\chi_{R1}\), this condition becomes
\begin{equation}
    \sin^2 2\chi_{R1}
    +2S_{R0,z}\sin\Delta\delta_1\sin2\chi_{R1}
    +S_{R0,z}^2-\cos^2\Delta\delta_1
    \leq0 .
    \label{eq:supp-3-1-1}
\end{equation}

We now consider a simple class of target transformations, where the distorted \(R\)-basis vector $\Vec{S}_{R3}$ is already equal to the recovered \(R\)-basis vector $\Vec{S}_{R0}=\Vec{S}_R=[0,0,1]^{\mathsf T}$. 
In this case, the target Mueller matrix leaves the \(R\)-axis of the Poincaré sphere unchanged and can only rotate the \(H\)-\(D\) plane. 
It has the form
\begin{equation}
    M_R(\phi)=
    \begin{bmatrix}
    \cos\phi & \sin\phi & 0\\
    -\sin\phi & \cos\phi & 0\\
    0 & 0 & 1
    \end{bmatrix}.
\end{equation}

For this family of targets, Eq.~\eqref{eq:supp-3-1-1} reduces to
\begin{equation}
    \left(\sin2\chi_{R1}+\sin\Delta\delta_1\right)^2\leq0.
\end{equation}
Therefore, a necessary condition is
\begin{equation}
    \sin2\chi_{R1}=-\sin\Delta\delta_1,
    \label{eq:supp-latitude-condition}
\end{equation}
or equivalently \(S_{R1,z}=-\sin\Delta\delta_1\). 
Thus, in the forward direction, a single non-ideal QWP with retardance \(\pi/2+\Delta\delta_1\) can map \(\vec S_R\) only onto this latitude circle on the Poincaré sphere. 
The azimuth on this circle can be chosen by the plate angle \(\theta_1\).

For example, using the angular branch \(2\chi_{R1}=-\Delta\delta_1\), Eq.~\eqref{eq:supp-sinB} gives \(\sin B=1\), so \(B=\pi/2\pmod{2\pi}\). 
Using the expression for \(S_{R1,z}\) obtained after the second QWP, Eq.~\eqref{eq:supp-latitude-condition} requires
\begin{equation}
    \cos D \sin\Delta\delta_3 \cos\Delta\delta_2
    +\cos\Delta\delta_3 \sin\Delta\delta_2
    =
    -\sin\Delta\delta_1 .
    \label{eq:supp-s18}
\end{equation}

In the ideal case, 
\(\Delta\delta_1=\Delta\delta_2=\Delta\delta_3=0\),
Eq.~\eqref{eq:supp-s18} reduces to \(0=0\), and hence does not restrict \(D\). 
This is consistent with the known result that two QWPs and one HWP with ideal retardances form a universal polarization gadget~\cite{Simon90,Bagini96}.

We next consider non-ideal retardances. 
If \(\Delta\delta_3=0\), then Eq.~\eqref{eq:supp-s18} reduces to
\begin{equation}
    \sin\Delta\delta_2=-\sin\Delta\delta_1 .
\end{equation}
For small retardance errors, this condition becomes
\begin{equation}
    \Delta\delta_1+\Delta\delta_2 = 0 .
\end{equation}
This condition is only satisfied in a lower-dimensional subset of the three-dimensional retardance error space. 
This should not be expected for a generic set of non-ideal wave plates in experiments. 
If \(\sin\Delta\delta_2\neq-\sin\Delta\delta_1\), then no value of \(D\) can satisfy Eq.~\eqref{eq:supp-s18}, therefore the family \(M_R(\phi)\) cannot be realized.

For the generic case
\(\sin\Delta\delta_3\cos\Delta\delta_2\neq0\), Eq.~\eqref{eq:supp-s18} gives
\begin{equation}
    \cos D =
    \frac{
    -\sin\Delta\delta_1-\cos\Delta\delta_3\sin\Delta\delta_2
    }{
    \sin\Delta\delta_3\cos\Delta\delta_2
    } .
    \label{eq:supp-cosD-condition}
\end{equation}
Therefore, a necessary condition for a real solution is
\begin{equation}
    \left|
    \sin\Delta\delta_1+\cos\Delta\delta_3\sin\Delta\delta_2
    \right|
    \leq
    \left|
    \sin\Delta\delta_3\cos\Delta\delta_2
    \right|.
    \label{eq:supp-D-existence-condition}
\end{equation}
When this condition is satisfied, Eq.~\eqref{eq:supp-cosD-condition} gives at most two values of \(D\) in a \(2\pi\) period. 
Thus, for generic non-ideal retardances, the relative angle between the second QWP and the HWP is not continuously adjustable but is restricted to a finite set.

For each admissible value of \(D\), and for each branch of \(B\), the remaining continuous freedom corresponds to a common rotation of all wave plate axes. 
If all three physical angles are shifted by \(\Delta\theta\), the total Mueller matrix is conjugated by an \(R\)-axis rotation,
\begin{equation}
    M_{\mathrm{QQH}}
    \mapsto
    M_R(2\Delta\theta)\,
    M_{\mathrm{QQH}}\,
    M_R(-2\Delta\theta).
\end{equation}
If \(M_{\mathrm{QQH}}=M_R(\phi)\), this conjugation leaves the matrix unchanged because rotations about the same axis commute:
\begin{equation}
    M_R(2\Delta\theta)M_R(\phi)M_R(-2\Delta\theta)
    =
    M_R(\phi).
\end{equation}
Therefore, the common rotation does not sweep out the continuous family \(M_R(\phi)\). 
Instead, each admissible branch produces at most one member of this subgroup. 
Since only finitely many branches and values of \(D\) are available, the non-ideal Q-Q-H sequence can realize at most finitely many matrices within the continuous \(R\)-axis rotation subgroup. 
Hence, it cannot realize all \(M_R(\phi)\), and therefore, it cannot realize arbitrary \(SO(3)\) Mueller matrices.

Thus, for generic non-ideal retardances, the Q-Q-H sequence fails to realize all members of the family \(M_R(\phi)\) of \(R\)-axis rotations. 
Since this family is only a subset of \(SO(3)\), the non-ideal Q-Q-H sequence cannot be universal over \(SO(3)\).

\subsection{Enhanced algorithm with an extra QWP}

As shown above, a sequence of three non-ideal wave plates cannot realize all \(SO(3)\) Mueller matrices in general. 
The Q-Q-Q-H angle-solving algorithm therefore introduces an additional QWP before the original Q-Q-H sequence, in order to remove the reachability obstruction introduced by non-ideal retardances. 
Denoting the Mueller matrix of this additional plate by \(M_+\), the total four-plate transformation is
\begin{equation}
    M_{\mathrm{QQQH}}=M_3M_2M_1M_+ .
\end{equation}
The role of \(M_+\) is to precondition the target vector before applying the Q-Q-H construction. 
In particular, the obstruction found above is demonstrated for the case \(S_{R0,z}=1\), where the target vector lies at the \(R\)-pole. 
We therefore choose the additional QWP so as to move \(\Vec{S}_{R0}\) close to the equatorial plane.

Let the actual retardance of the additional QWP be $\pi/2+\Delta\delta_+$, and let its slow-axis angle be \(\theta_+\). 
With the same Mueller matrix convention as above, its Mueller matrix is
\begin{equation}
M_+
=
\begin{bmatrix}
 \cos^2 2\theta_+ - \sin^2 2\theta_+ \sin\Delta\delta_+
&
 \cos2\theta_+ \sin2\theta_+ (1+\sin\Delta\delta_+)
&
 \sin2\theta_+ \cos\Delta\delta_+
\\
 \cos2\theta_+ \sin2\theta_+ (1+\sin\Delta\delta_+)
&
 -\cos^2 2\theta_+ \sin\Delta\delta_+ + \sin^2 2\theta_+
&
 -\cos2\theta_+ \cos\Delta\delta_+
\\
 -\sin2\theta_+ \cos\Delta\delta_+
&
 \cos2\theta_+ \cos\Delta\delta_+
&
 -\sin\Delta\delta_+
\end{bmatrix}.
\label{eq:supp-extra-qwp-matrix}
\end{equation}
The vector supplied to the remaining Q-Q-H sequence is
\begin{equation}
 \Vec{S}_{R0}'=M_+\Vec{S}_{R0}.
\end{equation}
For a target vector
\begin{equation}
 \Vec{S}_{R0}=
 [
 \cos2\psi_{R0}\cos2\chi_{R0},~
 \sin2\psi_{R0}\cos2\chi_{R0},~
 \sin2\chi_{R0}
 ]\T,
\end{equation}
we choose
\begin{equation}
 \theta_+=\psi_{R0}.
\end{equation}
The third component of \(M_+\Vec{S}_{R0}\) is
\begin{align}
S_{R0,z}'
=
\sin2\chi_{R0}'
&=
\sin2\theta_+ \cos\Delta\delta_+\, S_{R0,x}
-
\cos2\theta_+ \cos\Delta\delta_+\, S_{R0,y}
-
\sin\Delta\delta_+\, S_{R0,z} \nonumber\\
&=
-\sin\Delta\delta_+\,\sin2\chi_{R0}.
\label{eq:supp-extra-qwp-latitude}
\end{align}
Therefore
\begin{equation}
 \left|S_{R0,z}'\right|
 \leq
 \left|\sin\Delta\delta_+\right|.
\end{equation}
Equation~\eqref{eq:supp-extra-qwp-latitude} shows that the choice \(\theta_+=\psi_{R0}\) bounds the \(R\)-component of the vector supplied to the remaining Q-Q-H sequence:
\begin{equation}
 \left|S_{R0,z}'\right|
 \leq
 \left|\sin\Delta\delta_+\right| .
\end{equation}
Thus the additional QWP maps \(\Vec{S}_{R0}\) into a latitude band whose width is determined by its retardance error. 
In the ideal case this band collapses to the equator.

If \(\Vec{S}_{R0}\) lies at either pole of the Poincaré sphere, the transverse components of \(\Vec{S}_{R0}\) vanish, and Eq.~\eqref{eq:supp-extra-qwp-latitude} is independent of the azimuthal choice. 
Therefore \(\theta_+\) may be chosen arbitrarily in this polar case.

The choice \(\theta_+=\psi_{R0}\) is not the only possible choice. 
For a general value of \(\theta_+\), the same matrix gives
\begin{equation}
S_{R0,z}'
=
\cos\Delta\delta_+
\sin\!\left(2\theta_+-2\psi_{R0}\right)\cos2\chi_{R0}
-
\sin\Delta\delta_+\sin2\chi_{R0}.
\label{eq:supp-extra-qwp-general-latitude}
\end{equation}
Hence, choices of \(\theta_+\) that are not far from \(\psi_{R0}\) also keep the additional first term controlled. 
More generally, the additional QWP provides one extra continuous degree of freedom. 
Different values of \(\theta_+\) generate different preconditioned targets for the remaining Q-Q-H sequence, so the four-plate problem can have multiple solutions forming a low-dimensional continuous parameter space. 
In the following argument we nevertheless fix \(\theta_+=\psi_{R0}\).

We now derive a conservative condition under which the four-wave plate Q-Q-Q-H sequence is guaranteed to realize an arbitrary target matrix \(M\in SO(3)\). 
Since the additional QWP is placed before the original Q-Q-H sequence, fixing \(M_+\) defines the effective target matrix for the remaining three plates as
\begin{equation}
 M_{\mathrm{eff}} = M M_+^{-1}.
\end{equation}
Equivalently, the remaining Q-Q-H sequence should satisfy
\begin{equation}
 M_3M_2M_1 = M_{\mathrm{eff}} .
\end{equation}
The vector supplied to the Q-Q-H sequence is
\begin{equation}
    \Vec{S}_{R0}'=M_+\Vec{S}_{R0},
\end{equation}
and by construction
\begin{equation}
    M_{\mathrm{eff}}\Vec{S}_{R0}'=\Vec{S}_R.
\end{equation}

From Eq.~\eqref{eq:supp-psi-R1}, the azimuthal coordinate of \(\Vec{S}_{R1}\) can be adjusted by changing \(\theta_2\), since \(2\psi_{R1}\) contains the additive term \(2\theta_2\). 
The only exception is the singular case in which \(\Vec{S}_{R1}\) lies at a pole, where the azimuth is undefined but physically irrelevant. 
Therefore, the relevant constraint is in the latitude coordinate
\begin{equation}
    S_{R1,z}=\sin2\chi_{R1}.
\end{equation}
From Eq.~\eqref{eq:supp-chi-R1}, this coordinate is
\begin{equation}
    S_{R1,z}
    =
    \cos D\,\sin\Delta\delta_3\cos\Delta\delta_2
    +
    \cos\Delta\delta_3\sin\Delta\delta_2 .
    \label{eq:supp-SR1z-D}
\end{equation}
Hence, for all \(D\),
\begin{equation}
    |S_{R1,z}|
    \leq
    \sin|\Delta\delta_3|\cos|\Delta\delta_2|
    +
    \cos|\Delta\delta_3|\sin|\Delta\delta_2|.
\end{equation}
In the region $|\Delta\delta_2|+|\Delta\delta_3|<\pi/2$, this gives the simpler bound
\begin{equation}
    |S_{R1,z}|
    \leq
    \sin\left(|\Delta\delta_2|+|\Delta\delta_3|\right).
    \label{eq:supp-SR1z-bound}
\end{equation}

Using \(S_{R0,z}'\) in place of \(S_{R0,z}\), Eq.~\eqref{eq:supp-3-1-1} becomes
\[
 \left(\sin2\chi_{R1}\right)^2
 +2S_{R0,z}'\sin\Delta\delta_1\,\sin2\chi_{R1}
 +(S_{R0,z}')^2-\cos^2\Delta\delta_1
 \leq0 .
\]
Therefore, a real solution for \(B\) exists if and only if
\begin{equation}
 \sin2\chi_{R1}
 \in
 \left[
 -S_{R0,z}'\sin\Delta\delta_1
 -
 \left|\cos\Delta\delta_1\right|
 \sqrt{1-(S_{R0,z}')^2},
 \,
 -S_{R0,z}'\sin\Delta\delta_1
 +
 \left|\cos\Delta\delta_1\right|
 \sqrt{1-(S_{R0,z}')^2}
 \right].
 \label{eq:supp-B-existence-interval}
\end{equation}

From the choice of the additional QWP in Eq.~\eqref{eq:supp-extra-qwp-latitude}, we have
\begin{equation}
    |S_{R0,z}'|\leq|\sin\Delta\delta_+|.
\end{equation}
Thus
\begin{equation}
    \sqrt{1-(S_{R0,z}')^2}
    \geq
    |\cos\Delta\delta_+|.
\end{equation}
Moreover, the center of the interval in Eq.~\eqref{eq:supp-B-existence-interval} can shift by at most $|\sin\Delta\delta_+\,\sin\Delta\delta_1|$.
It is therefore sufficient that every reachable value of \(S_{R1,z}\) satisfies
\begin{equation}
    |S_{R1,z}|
    +
    |\sin\Delta\delta_+\,\sin\Delta\delta_1|
    \leq
    |\cos\Delta\delta_+\,\cos\Delta\delta_1|.
    \label{eq:supp-conservative-B-condition}
\end{equation}

We now give a simple sufficient condition for Eq.~\eqref{eq:supp-conservative-B-condition}. 
Define
\begin{equation}
    p=|\Delta\delta_+|,
    \qquad
    q=|\Delta\delta_1|,
    \qquad
    r=|\Delta\delta_2|+|\Delta\delta_3|.
\end{equation}
Assume that
\begin{equation}
    p+q+r
    =
    |\Delta\delta_+|
    +
    |\Delta\delta_1|
    +
    |\Delta\delta_2|
    +
    |\Delta\delta_3|
    <
    \frac{\pi}{2}.
    \label{eq:supp-total-error-bound}
\end{equation}
Then \(r<\pi/2-p-q\), and Eq.~\eqref{eq:supp-SR1z-bound} gives
\begin{equation}
    |S_{R1,z}|\leq\sin r.
\end{equation}
Since \(r<\pi/2-p-q\), we have
\begin{equation}
    \sin r < \cos(p+q).
\end{equation}
Using
\begin{equation}
    \cos(p+q)=\cos p\cos q-\sin p\sin q,
\end{equation}
we obtain
\begin{equation}
    |S_{R1,z}|
    +
    |\sin\Delta\delta_+\,\sin\Delta\delta_1|
    <
    \cos p\cos q.
\end{equation}
Since \(p,q<\pi/2\), this is exactly Eq.~\eqref{eq:supp-conservative-B-condition}. 
Therefore, under the bound from Eq.~\eqref{eq:supp-total-error-bound}, a branch of $B$ exists for every value of \(D\). 
In particular, for every \(D\), the construction can satisfy the required condition on the \(R\)-basis vector. 

It remains to show that the components perpendicular to the \(R\)-axis can be rotated to the required values. 
For each admissible value of \(D\), let
\begin{equation}
    Q_b(D)=M_3M_2M_1
\end{equation}
be the Q-Q-H matrix obtained from a continuous branch \(b\) of the
solution of $B$. 
By the construction above,
\[
    Q_b(D)\Vec{S}_{R0}'=\Vec{S}_R,
    \qquad
    M_{\mathrm{eff}}\Vec{S}_{R0}'=\Vec{S}_R.
\]
Therefore, the residual matrix
\begin{equation}
    R_b(D)=Q_b(D)M_{\mathrm{eff}}^{-1}
    \label{eq:supp-residual-matrix}
\end{equation}
leaves \(\Vec{S}_R\) invariant. 
Since \(R_b(D)\in SO(3)\), it must be a rotation about the \(R\)-axis. 
We write this rotation as
\begin{equation}
 R_b(D)=M_R(\Phi_b(D)),
 \label{eq:supp-residual-angle}
\end{equation}
where \(\Phi_b(D)\) is the residual rotation angle around the 
\(R\)-axis for branch \(b\). 
The full Mueller matrix condition \(Q_b(D)=M_{\mathrm{eff}}\) is therefore equivalent to
\begin{equation}
 \Phi_b(D)=0
 \pmod{2\pi}.
 \label{eq:supp-residual-zero}
\end{equation}

We first evaluate the winding number of this residual angle in the ideal retardance limit. 
Set
\begin{equation}
 \Delta\delta_+
 =
 \Delta\delta_1
 =
 \Delta\delta_2
 =
 \Delta\delta_3
 =
 0 .
\end{equation}
After the additional ideal QWP, the vector \(\Vec{S}_{R0}'\) lies on the equator. 
Write
\begin{equation}
 \Vec{S}_{R0}'=
 [
 \cos2\psi_0,~
 \sin2\psi_0,~
 0
 ]\T.
\end{equation}
For the \(B=0\) branch of the ideal Q-Q-H construction, the angles are
\begin{equation}
 \theta_1=\psi_0,\qquad
 \theta_2=\psi_0+\frac{\pi}{4},\qquad
 \theta_3=\psi_0+\frac{\pi}{4}+\frac{D}{2}.
 \label{eq:supp-ideal-branch-angles}
\end{equation}
The winding number is independent of the absolute azimuth \(\psi_0\), so we set \(\psi_0=0\) for simplicity. 
Then
\begin{equation}
 \theta_1=0,\qquad
 \theta_2=\frac{\pi}{4},\qquad
 \theta_3=\frac{\pi}{4}+\frac{D}{2}.
\end{equation}
The resulting ideal Q-Q-H Mueller matrix is
\begin{equation}
 Q_0(D)=
 \begin{bmatrix}
 0 & -\cos2D & \sin2D\\
 0 & -\sin2D & -\cos2D\\
 1 & 0 & 0
 \end{bmatrix}.
 \label{eq:supp-ideal-QD}
\end{equation}
Thus \(Q_0(D)\Vec{S}_H=\Vec{S}_R\) for all \(D\). 
In particular, \(Q_0(0)\) and \(M_{\mathrm{eff}}\) both satisfy the same \(R\)-basis vector condition, 
\begin{equation}
 Q_0(0)\Vec{S}_{R0}'=\Vec{S}_R,
 \qquad
 M_{\mathrm{eff}}\Vec{S}_{R0}'=\Vec{S}_R .
\end{equation}
It follows that
\begin{equation}
 Q_0(0)M_{\mathrm{eff}}^{-1}\Vec{S}_R=\Vec{S}_R .
\end{equation}
Since \(Q_0(0)M_{\mathrm{eff}}^{-1}\in SO(3)\), it is a rotation about the \(R\)-axis. 
Therefore there exists a constant angle \(\Phi_0\), defined modulo \(2\pi\), such that
\begin{equation}
 Q_0(0)M_{\mathrm{eff}}^{-1}=M_R(\Phi_0).
\end{equation}

Using \(Q_0(0)\) as a reference, we have
\begin{equation}
 Q_0(D)Q_0(0)^{\mathsf T}
 =
 \begin{bmatrix}
 \cos2D & -\sin2D & 0\\
 \sin2D & \cos2D & 0\\
 0 & 0 & 1
 \end{bmatrix}
 =
 M_R(-2D).
 \label{eq:supp-ideal-residual}
\end{equation}
Therefore, the ideal residual matrix for the effective target is
\begin{equation}
 Q_0(D)M_{\mathrm{eff}}^{-1}
 =
 Q_0(D)Q_0(0)^{\mathsf T}
 Q_0(0)M_{\mathrm{eff}}^{-1}
 =
 M_R(-2D)M_R(\Phi_0)
 =
 M_R(\Phi_0-2D).
\end{equation}
Thus the residual angle \(\Phi_b(D)\), which is defined only modulo \(2\pi\), has in this ideal branch a continuous lift
\begin{equation}
 \widetilde{\Phi}_0(D)=\Phi_0-2D ,
\end{equation}
where the tilde denotes the chosen continuous version of the angle, rather than its value modulo \(2\pi\). 
Consequently,
\begin{equation}
 \widetilde{\Phi}_0(D+2\pi)-\widetilde{\Phi}_0(D)
 =
 -4\pi,
\end{equation}
and the winding number is
\begin{equation}
 w_0
 =
 \frac{-4\pi}{2\pi}
 =
 -2 .
 \label{eq:supp-ideal-winding}
\end{equation}

For an arbitrary effective target matrix \(M_{\mathrm{eff}}\), the ideal residual angle is shifted only by a constant. Since both \(Q_0(0)\) and \(M_{\mathrm{eff}}\) map \(\Vec{S}_{R0}'\) to \(\Vec{S}_R\), the matrix $Q_0(0)M_{\mathrm{eff}}^{-1}$ also leaves \(\Vec{S}_R\) invariant and therefore equals \(M_R(\Phi_0)\) for some constant \(\Phi_0\). 
Consequently,
\begin{equation}
    Q_0(D)M_{\mathrm{eff}}^{-1}
    =
    Q_0(D)Q_0(0)^{\mathsf T}Q_0(0)M_{\mathrm{eff}}^{-1}
    =
    M_R(-2D)M_R(\Phi_0)
    =
    M_R(\Phi_0-2D).
\end{equation}
Thus, the winding number remains \(-2\), independent of the target matrix.

We now return to non-ideal retardances satisfying Eq.~\eqref{eq:supp-total-error-bound}. 
To connect the ideal and non-ideal cases, we have to scale the retardance errors continuously as
\begin{equation}
 \Delta\delta_i(\tau)=\tau\Delta\delta_i,
 \qquad
 \tau\in[0,1],
 \qquad
 i\in\{+,1,2,3\}.
\end{equation}
Here \(\tau=0\) gives the ideal retardance case, while \(\tau=1\) gives the actual non-ideal wave plates. 
Because the strict bound \eqref{eq:supp-total-error-bound} holds at \(\tau=1\), it also holds for every \(\tau\in[0,1]\). 
Therefore, the \(B\)-branch remains non-singular during this continuous change of the retardance errors.

For each \(\tau\), let \(\widetilde{\Phi}_b(D;\tau)\) be a continuous lift of the residual angle on the same branch. 
The change of this lifted angle over one \(2\pi\) period of \(D\), $\widetilde{\Phi}_b(D+2\pi;\tau)-\widetilde{\Phi}_b(D;\tau)$, varies continuously with \(\tau\). 
At the same time, this change must be an integer multiple of \(2\pi\), because \(\Phi_b(D;\tau)\) itself is defined modulo \(2\pi\). 
A quantity that is both continuous in \(\tau\) and restricted to integer multiples of \(2\pi\) cannot change unless a singularity occurs. 
Since the branch remains non-singular, the value found in the ideal case is preserved:
\begin{equation}
 \widetilde{\Phi}_b(D+2\pi;1)-\widetilde{\Phi}_b(D;1)
 =
 -4\pi .
\end{equation}
Equivalently, the non-ideal branch has the same winding number as the ideal branch,
\begin{equation}
 w_b=-2 .
\end{equation}
Consequently, as \(D\) varies over one \(2\pi\) period, the residual angle passes through every value modulo \(2\pi\). 
In particular, there exists at least one value of \(D\) for which
\begin{equation}
 \Phi_b(D)=0\pmod{2\pi}.
\end{equation}

For this value of \(D\), Eq.~\eqref{eq:supp-residual-zero} holds, so
\begin{equation}
    Q_b(D)=M_{\mathrm{eff}}.
\end{equation}
Therefore
\begin{equation}
    M_3M_2M_1M_+=M.
\end{equation}

We have thus shown that, under the conservative condition
\begin{equation}
    |\Delta\delta_+|
    +
    |\Delta\delta_1|
    +
    |\Delta\delta_2|
    +
    |\Delta\delta_3|
    <
    \frac{\pi}{2},
    \label{eq:supp-conservative_condition}
\end{equation}
the non-ideal Q-Q-Q-H sequence can realize an arbitrary \(SO(3)\) Mueller matrix. 
The additional QWP therefore removes the reachability obstruction present in the non-ideal Q-Q-H sequence.

\subsection{Retardance error budget of the commercial wave plates}

We evaluate the sufficient condition derived above using the retardance data for the commercial wave plates provided by the manufacturers. 
Using linear interpolation of the datasheet values, Eq.~\eqref{eq:supp-conservative_condition} is satisfied from \(535.4\,\mathrm{nm}\) to at least \(750\,\mathrm{nm}\) for the zero-order \(633\,\mathrm{nm}\) plates WPQ05M-633 and WPH05M-633.
The upper bound is written as ``at least'' because the available HWP data extend only to \(750\,\mathrm{nm}\), where the condition is still satisfied. 
For the corresponding multi-order \(633\,\mathrm{nm}\) plates, the same calculation gives a much narrower range of \SI{629.7}{nm} to \SI{634.7}{nm}. 
The zero-order plates provide a substantially larger guaranteed wavelength range for arbitrary \(SO(3)\) compensation. 

Because Eq.~\eqref{eq:supp-conservative_condition} is only a sufficient condition, failure to satisfy it does not necessarily imply loss of reachability. 
We therefore tested the \(515\,\mathrm{nm}\) case numerically by generating target Mueller matrices from the three-parameter form
\begin{equation}
    M =
    \begin{bmatrix}
        \cos(2\alpha) \cos(2\gamma) + \sin(2\alpha) \sin(2\gamma) \cos\delta
        &
        \cos(2\alpha) \sin(2\gamma) - \sin(2\alpha) \cos(2\gamma) \cos\delta
        &
        \sin(2\alpha) \sin\delta
        \\
        \sin(2\alpha) \cos(2\gamma) - \cos(2\alpha) \sin(2\gamma) \cos\delta
        &
        \sin(2\alpha) \sin(2\gamma) + \cos(2\alpha) \cos(2\gamma) \cos\delta
        &
        -\cos(2\alpha) \sin\delta
        \\
        -\sin(2\gamma) \sin\delta
        &
        \cos(2\gamma) \sin\delta
        &
        \cos\delta
    \end{bmatrix}.
    \label{eq:supp-so3-grid-parametrization}
\end{equation}
Here \(\alpha\), \(\gamma\), and \(\delta\) are the angular parameters used to sample \(SO(3)\). 
The symbol \(\delta\) in this equation is not necessarily a wave plate retardance value. 
For the zero-order \(633\,\mathrm{nm}\) plates, the Q-Q-Q-H solver found at least one solution branch for every tested target matrix at \(515\,\mathrm{nm}\), although the sufficient condition is not fulfilled at this wavelength. 
This numerical grid does not serve as a proof of full reachability at \(\SI{515}{nm}\), but indicates that the soluble wavelength range of the zero-order Q-Q-Q-H compensator is larger than the conservative guarantee in Eq.~\eqref{eq:supp-conservative_condition}. 

\section{Measuring Mueller matrices with eight Stokes vectors}

As discussed in the main text, the wave plate compensator may be inserted inside the optical link rather than placed immediately before the detector. 
In this case, the polarization transformation before the compensator is denoted by \(M_\alpha\), and the transformation after the compensator is denoted by \(M_\beta\). 
Under the same loss-normalized, non-depolarizing assumptions as above, all polarization transformations are represented by matrices in \(SO(3)\).

For an input normalized Stokes vector \(\vec S_{\rm in}\), the output Stokes vector is therefore
\begin{equation}
    \vec S_{\rm out}
    =
    M_\beta M M_\alpha \vec S_{\rm in},
\end{equation}
where \(M\) is the Mueller matrix implemented by the wave plate stack. 
Because matrix multiplication is not commutative, the required middle-link compensation matrix is not the inverse of the end-to-end uncompensated transformation \(M_\beta M_\alpha\) in general.
Instead, perfect compensation requires
\begin{equation}
    M_\beta M M_\alpha = I .
\end{equation}
Solving for \(M\) gives
\begin{equation}
    M = M_\beta^{-1}M_\alpha^{-1}.
\end{equation}
Since \(M_\alpha,M_\beta\in SO(3)\), their inverses are equal to their transposes, and hence
\begin{equation}
    M = M_\beta^{\mathsf T}M_\alpha^{\mathsf T}.
\end{equation}
Thus, the two channel segments must be characterized separately when the compensator is placed inside the link; the uncompensated product \(M_\beta M_\alpha\) alone does not determine the required compensating transformation.
Instead, using the inverse of the uncompensated end-to-end matrix would give $M=(M_\beta M_\alpha)^{-1}=M_\alpha^{-1}M_\beta^{-1}$,
which yields $M_\beta M M_\alpha = M_\beta M_\alpha^{-1}M_\beta^{-1}M_\alpha$, not \(I\) unless special commutation conditions are satisfied. 

In this section, we first show how a set of known wave plate transformations can be used to reconstruct the input- and output-side link matrices \(M_\alpha\) and \(M_\beta\). 
We then describe the processing of non-ideal experimental data, including how the measured Stokes vector pairs are orthonormalized before building the end-to-end Mueller matrices, and how the  reconstructed \(M_\beta\) is restricted onto \(SO(3)\).

\subsection{Mueller matrix reconstruction from test rotations}

If two wave plate Mueller matrices \(M'\) and \(M''\) are chosen, the corresponding end-to-end transformations $M_\beta M'M_\alpha$ and $M_\beta M''M_\alpha$ can be obtained by setting the wave plates to the corresponding positions and measuring the distorted Stokes vectors for the input states \(H\) and \(D\).
Since \(M_\beta M M_\alpha\in SO(3)\), the third column is fixed by the first two columns. 
Therefore,
\begin{equation}
    M_\beta M M_\alpha =
    \begin{bmatrix}
        M_\beta M M_\alpha \vec S_H &
        M_\beta M M_\alpha \vec S_D &
        \left(M_\beta M M_\alpha \vec S_H\right)
        \times
        \left(M_\beta M M_\alpha \vec S_D\right)
    \end{bmatrix}.
\end{equation}

Multiplying one measured matrix by the transpose of the other eliminates \(M_\alpha\):
\begin{align}
    \left(M_\beta M'M_\alpha\right)
    \left(M_\beta M''M_\alpha\right)^{\mathsf T}
    &=
    M_\beta M'M_\alpha
    M_\alpha^{\mathsf T}
    (M'')^{\mathsf T}
    M_\beta^{\mathsf T} \\
    &=
    M_\beta M'(M'')^{\mathsf T}M_\beta^{\mathsf T}.
\end{align}

We define
\begin{equation}
    N=M'(M'')^{\mathsf T}.
\end{equation}
Assuming that the wave plate stack can realize arbitrary elements of \(SO(3)\), \(N\) can be chosen freely. 
For example, one may choose \(M'=N\) and \(M''=I\). 
We choose three test rotations about the Cartesian Stokes axes by an angle \(0<\Theta<\pi\):
    \begin{align}
        N_1 &=
        \begin{bmatrix}
            1 & 0 & 0 \\
            0 & \cos\Theta & -\sin\Theta \\
            0 & \sin\Theta & \cos\Theta
        \end{bmatrix}, \nonumber \\
        N_2 &=
        \begin{bmatrix}
            \cos\Theta & 0 & \sin\Theta \\
            0 & 1 & 0 \\
            -\sin\Theta & 0 & \cos\Theta
        \end{bmatrix}, \nonumber \\
        N_3 &=
        \begin{bmatrix}
            \cos\Theta & -\sin\Theta & 0 \\
            \sin\Theta & \cos\Theta & 0 \\
            0 & 0 & 1
        \end{bmatrix}.
    \end{align}

For each \(i\), we denote
\begin{equation}
    C_i=M_\beta N_iM_\beta^{\mathsf T}.
\end{equation}
Since \(C_i\) is obtained from \(N_i\) by conjugation with \(M_\beta\), it is also a rotation by the angle \(\Theta\). 
Its rotation axis is the corresponding Cartesian Stokes axis multiplied with \(M_\beta\). 
Thus, with
\begin{equation}
    \hat e_1=\vec S_H,\qquad
    \hat e_2=\vec S_D,\qquad
    \hat e_3=\vec S_R,
\end{equation}
the rotation axis of \(C_i\) is
\begin{equation}
    \vec u_i=M_\beta \hat e_i .
\end{equation}
Therefore,
\begin{equation}
    M_\beta=
    \begin{bmatrix}
        \vec u_1 & \vec u_2 & \vec u_3
    \end{bmatrix}.
\end{equation}

To extract the rotation axis \(\vec u_i\) from \(C_i\), we use the antisymmetric part of a rotation matrix. 
Consider a general rotation by angle \(\Theta\) about a unit axis \(\vec u\), and denote the corresponding rotation matrix by \(C\). 
For an arbitrary vector \(\vec a\), let
\begin{equation}
    \vec a_+ = C\vec a,
    \qquad
    \vec a_- = C^{\mathsf T}\vec a,
\end{equation}
where \(C^{\mathsf T}\) represents the inverse rotation, i.e., a rotation by \(-\Theta\) about the same axis. 
From the geometry of rotations in a right-handed frame, one obtains
\begin{equation}
    \vec a_+ - \vec a_-
    =
    2\sin\Theta\, \vec u \times \vec a .
    \label{eq:supp-s32}
\end{equation}

We now take \(\vec a=\hat{x}\). 
Then \(C\hat{x}\) is the first column of \(C\), while \(C^{\mathsf T}\hat{x}\) is the transpose of the first row of \(C\). 
Substituting \(\vec a=\hat{x}\) into Eq.~\eqref{eq:supp-s32} gives
\begin{equation}
    C\hat{x}-C^{\mathsf T}\hat{x}
    =
    2\sin\Theta
    \begin{bmatrix}
        0\\
        u_z\\
        -u_y
    \end{bmatrix}.
\end{equation}
Therefore,
\begin{equation}
    u_z=\frac{C_{yx}-C_{xy}}{2\sin\Theta},
    \qquad
    u_y=\frac{C_{xz}-C_{zx}}{2\sin\Theta}.
\end{equation}
Applying this to the other Cartesian basis vectors gives
\begin{equation}
    u_x=\frac{C_{zy}-C_{yz}}{2\sin\Theta}.
\end{equation}

Therefore, for each measured matrix \(C_i\), the corresponding rotation axis is obtained as
\begin{equation}
    \vec u_i
    =
    \frac{1}{2\sin\Theta}
    \begin{bmatrix}
        (C_i)_{zy}-(C_i)_{yz} \\
        (C_i)_{xz}-(C_i)_{zx} \\
        (C_i)_{yx}-(C_i)_{xy}
    \end{bmatrix}.
    \label{eq:supp-s33}
\end{equation}

In the experiment, we choose \(\Theta=\pi/2\). 
This choice maximizes \(\sin\Theta\), and therefore maximizes the magnitude of the antisymmetric matrix elements in Eq.~\eqref{eq:supp-s33}. 
It consequently improves the signal-to-noise ratio of the extracted rotation axis and reduces the sensitivity to measurement errors.

\subsection{Orthonormalization of measured Stokes vector pairs}
\label{sec:supp-stokes-orthonormalization}

For each wave plate setting \(N_i\), the measured output Stokes vectors for the two input states \(H\) and \(D\) are denoted by \(\vec a_i\) and \(\vec b_i\), respectively. 
In the ideal \(SO(3)\) model these vectors are unit length and mutually orthogonal, and the corresponding end-to-end Mueller matrix is
\begin{equation}
    E_i
    =
    \begin{bmatrix}
        \vec a_i & \vec b_i & \vec a_i\times \vec b_i
    \end{bmatrix}.
    \label{eq:supp-end-to-end-matrix}
\end{equation}
In the experiments, finite polarimeter accuracy and other measurement errors can make the measured pair slightly non-orthogonal. 
We therefore replace the pair by an orthonormal pair while preserving the angular bisector. 
Assuming \(\vec a_i\) and \(\vec b_i\) have been normalized, define
\begin{equation}
    \vec m_i=
    \frac{\vec a_i+\vec b_i}{\|\vec a_i+\vec b_i\|},
    \qquad
    \vec d_i=
    \frac{\vec a_i-\vec b_i}{\|\vec a_i-\vec b_i\|}.
    \label{eq:supp-bisector-difference}
\end{equation}
The orthonormalized vectors are
\begin{equation}
    \tilde a_i=
    \frac{\vec m_i+\vec d_i}{\sqrt{2}},
    \qquad
    \tilde b_i=
    \frac{\vec m_i-\vec d_i}{\sqrt{2}}.
    \label{eq:supp-orthogonalized-HD}
\end{equation}
They satisfy \(\tilde a_i\cdot\tilde b_i=0\), and the matrix used in the reconstruction is
\begin{equation}
    E_i
    =
    \begin{bmatrix}
        \tilde a_i & \tilde b_i & \tilde a_i\times\tilde b_i
    \end{bmatrix}.
    \label{eq:supp-orthogonalized-end-to-end}
\end{equation}

\subsection{Orthonormalization of the reconstructed output-side link matrix}
\label{sec:supp-Mbeta-orthonormalization}

For the three test rotations \(N_i\), the extracted rotation axes are denoted by \(\vec u_i\). 
In the ideal \(SO(3)\) model, these vectors are unit length, mutually orthogonal, and form the columns of the output-side link matrix,
\begin{equation}
    U_\beta
    =
    \begin{bmatrix}
        \vec u_1 & \vec u_2 & \vec u_3
    \end{bmatrix}
    =
    M_\beta .
\end{equation}
In the experiments, imperfect realization of the test rotations and measurement errors can lead to a slightly non-orthonormal \(U_\beta\). 
We hence project \(U_\beta\) onto the nearest proper rotation matrix before using it as \(M_\beta\).

Let
\begin{equation}
    U_\beta = A\Sigma B^{\mathsf T}
\end{equation}
be the singular value decomposition (SVD) of \(U_\beta\). 
The orthonormalized output-side link matrix is calculated as
\begin{equation}
    M_\beta
    =
    A
    \begin{bmatrix}
        1 & 0 & 0\\
        0 & 1 & 0\\
        0 & 0 & \det(A B^{\mathsf T})
    \end{bmatrix}
    B^{\mathsf T}.
    \label{eq:supp-Mbeta-svd-projection}
\end{equation}
This gives \(M_\beta^{\mathsf T}M_\beta=I\) and \(\det M_\beta=1\). 
The input-side link matrix is then
\begin{equation}
    M_\alpha=M_\beta^{\mathsf T}E_0 .
\end{equation}
Thus both estimated link-segment matrices are constrained to lie in \(SO(3)\). 

\section{Experimental hardware characterization}

\subsection{Laser spectra}

\begin{figure}
    \centering

    \begin{minipage}[t]{0.48\linewidth}
        \centering
        \textbf{(a)}\\[-0.5ex]
        \includegraphics[width=\linewidth]{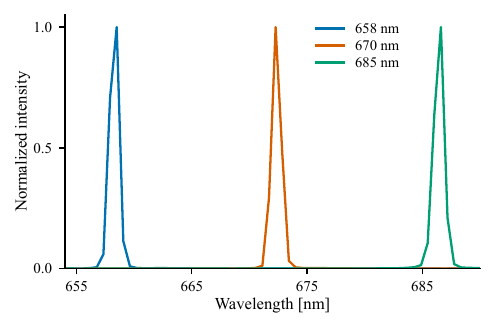}
    \end{minipage}
    \hfill
    \begin{minipage}[t]{0.48\linewidth}
        \centering
        \textbf{(b)}\\[-0.5ex]
        \includegraphics[width=\linewidth]{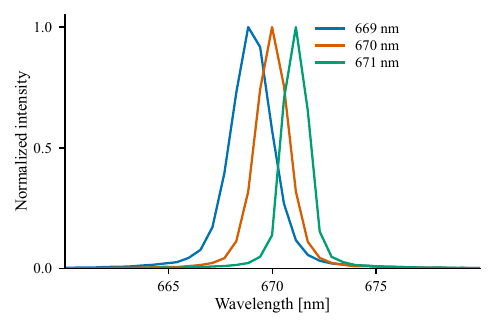}
    \end{minipage}

    \caption{
    Diode laser spectra used in the experiments.
    \textbf{(a)} Spectra of the diode lasers with nominal wavelengths of \SI{658}{nm}, \SI{670}{nm}, and \SI{685}{nm}. 
    Each spectrum is normalized to its own maximum. 
    The measured peak wavelengths are \SI{658.3}{nm}, \SI{672.3}{nm}, and \SI{686.5}{nm}, respectively.
    \textbf{(b)} Spectra obtained by current tuning of the nominal \SI{670}{nm} diode for the close-wavelength auxiliary experiment, with emission peaks near \SI{669}{nm}, \SI{670}{nm}, and \SI{671}{nm}. 
    }
    \label{fig:supp_laser_spectra}
\end{figure}

Most experiments were performed using diode lasers as optical sources. 
The lasers used for the auxiliary  and signal wavelength measurements were L658P040, HL6748MG, and HL6750MG diodes, with nominal center wavelengths of \SI{658}{nm}, \SI{670}{nm}, and \SI{685}{nm}, respectively. 
The measured optical spectra are shown in Fig.~\ref{fig:supp_laser_spectra}(a), where each spectrum is normalized to its own maximum. 
The corresponding peak wavelengths were \SI{658.3}{nm}, \SI{672.3}{nm}, and \SI{686.5}{nm}, respectively.

For the close-wavelength auxiliary compensation experiment, in which the two auxiliary wavelengths were separated by approximately \SI{2}{nm}, the emission wavelength of the nominal \SI{670}{nm} diode was adjusted by changing its drive current. 
The resulting spectra are shown in Fig.~\ref{fig:supp_laser_spectra}(b). 
This current tuning provided the closely spaced wavelengths used for the \SI{669}{nm}/\SI{670}{nm}/\SI{671}{nm} configuration discussed in the main text. 

The finite linewidth of the current-tuned diode may reduce the effective degree of polarization, because different spectral components can experience slightly different polarization transformations. 
This effect is not a fundamental limitation of the method and can be mitigated in practical implementations by using sources with narrower linewidths, together with standard wavelength-division multiplexed (WDM) devices. 

\subsection{Baseline performance and hardware limitations}

\begin{figure}
    \centering
    \includegraphics[width=0.95\textwidth]{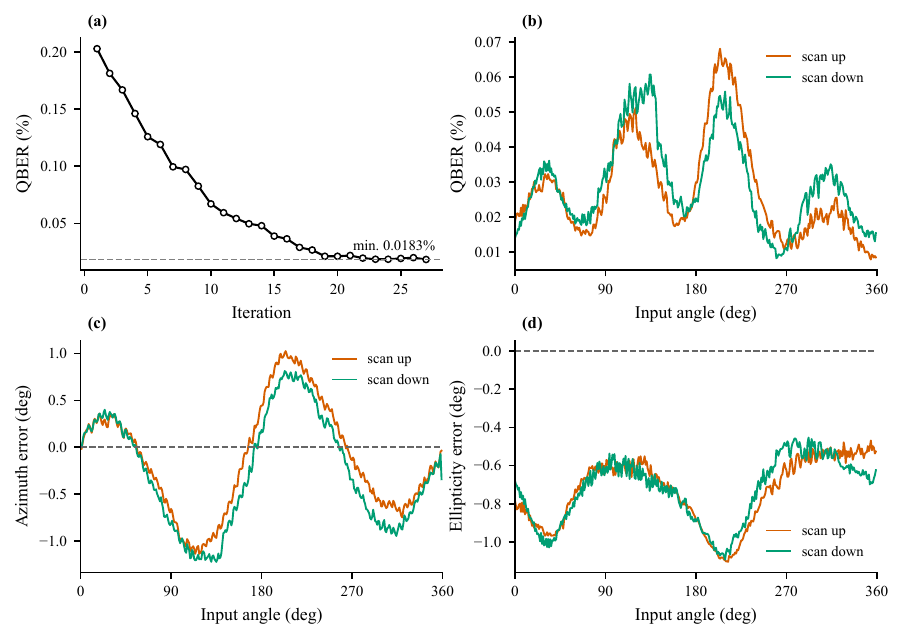}
    \caption{
    Baseline characterization at the signal wavelength, \(\lambda=\SI{672.3}{nm}\), without auxiliary wavelength interpolation. 
    \textbf{(a)} Direct local optimization of the polarization-induced excess QBER by adjusting the wave plate angles at the signal wavelength. 
    \textbf{(b)} Polarization-induced excess QBER estimated from an independent scan of input linear polarization angles. 
    \textbf{(c)} Apparent azimuth error extracted from upward and downward angular scans.
    \textbf{(d)} Ellipticity offset from the nominal linear polarization condition. 
    Panels (b)--(d) were derived from the same scan, which was not performed after fully minimizing the QBER to the same local optimum as in panel (a). 
    }
    \label{fig:supp_baseline_limitations_stokes_qber}
\end{figure}

To estimate the baseline performance of the experimental setup, we first performed a direct local optimization of the polarization-induced excess quantum bit error rate (QBER) at the signal wavelength, \(\lambda=\SI{672.3}{nm}\). 
In this test, the wave plate angles were adjusted directly using the measured polarization error at the signal wavelength. 
As shown in Fig.~\ref{fig:supp_baseline_limitations_stokes_qber}(a), the QBER decreased during the optimization. 
In the final part of this run, after 19 iterations, the optimization no longer changed the wave plate settings in subsequent iterations. 
The residual QBER at this stationary setting remained around \(\SI{0.019}{\percent}\), with the lowest sampled value being \(\SI{0.0183}{\percent}\). 

This value serves as an experimental baseline for the present setup, but does not reflect the fundamental limit of the compensation method. 
At this level, several nonidealities in state preparation, polarization analysis, and wave plate positioning can contribute comparable residual errors. 
To characterize these effects independently, we also scanned the input linear polarization angle and reconstructed the corresponding output Stokes vectors. 
The QBER values shown in Fig.~\ref{fig:supp_baseline_limitations_stokes_qber}(b) were calculated from the overlap between the measured Stokes vector and the target Stokes vector,
\begin{equation}
    {\rm QBER}_{\rm pol}
    =
    \frac{1-\vec S\cdot\vec S'}{2}.
\end{equation}
Note that this scan was not performed after fully minimizing the QBER to the same local optimum as in Fig.~\ref{fig:supp_baseline_limitations_stokes_qber}(a).

The same data were used to estimate the apparent azimuth error and ellipticity offset of the prepared states. 
The azimuth error was obtained by comparing the measured output azimuth with the input angle, while the ellipticity offset quantified the deviation from the nominal linear polarization condition. 
The maximum apparent azimuth error in this scan was approximately \(\SI{1.2}{\degree}\). 
Because this estimate is inferred from the Stokes vector measurement, it can include imperfections in state preparation, analysis, and actuation, such as the finite angular accuracy and backlash of the motorized rotation mount, as well as alignment, polarizer, and polarimeter calibration errors. 

The finite extinction ratio of the input polarizer provides another possible contribution to the baseline error. 
The Thorlabs WP25M-VIS wire-grid polarizer has an extinction ratio specified to be better than \(683{:}1\). 
As an estimate, consider a circularly polarized state incident on a linear polarizer whose intensity transmission ratio for the two orthogonal linear components is \(683{:}1\). 
After normalization, the output Stokes vector is
\begin{equation}
    \vec S
    =
    \begin{bmatrix}
    \dfrac{683-1}{683+1} \\
    0 \\
    -\dfrac{2\sqrt{683}}{683+1}
    \end{bmatrix}
    \simeq
    \begin{bmatrix}
    0.997 \\
    0 \\
    -0.076
    \end{bmatrix}.
    \nonumber
\end{equation}
Its overlap with the ideal horizontal state is therefore
\((683-1)/(683+1)\), corresponding to
\begin{equation}
    {\rm QBER}_{\rm ext}
    =
    \frac{1}{2}
    \left(
    1-\frac{683-1}{683+1}
    \right)
    =
    \frac{1}{684}
    \simeq
    \SI{0.146}{\percent}.
    \nonumber
\end{equation}
This calculation is not a direct measurement of the residual QBER in the experiment. 
Rather, it gives an estimate showing that finite state preparation extinction alone could produce a polarization mismatch nearly one order of magnitude larger than the optimized residual QBER observed in Fig.~\ref{fig:supp_baseline_limitations_stokes_qber}(a).

\section{Axis-angle representation of polarization transformations}

Although Mueller matrices are the primary representation used in this work, the relevant information is not always easy to read directly from the matrix elements themselves. 
We therefore also use an equivalent axis-angle parametrization for each loss-normalized non-depolarizing polarization transformation. 
This representation provides a locally continuous coordinate system for interpolating auxiliary wavelength measurements, and also gives a compact way to visualize the measured spectral or thermal drift of a Mueller matrix. 
We summarize here the conversion between the three-component Mueller matrix, the corresponding Jones representative, and the axis-angle parameters. 

\subsection{Mueller and Jones representatives}

The \(SO(3)\) Mueller matrices used in this work can also be represented by Jones matrices. 
This section gives a Jones parametrization used for interpolation and the conversion formulas between this representation and the corresponding \(SO(3)\) Mueller matrix \cite{Chartier01}.

\subsubsection{Mueller to Jones}

Let \(M\in SO(3)\) be the three-component Mueller matrix. 
We use the phase-fixed Jones representative
\begin{equation}
J=
\begin{bmatrix}
a & b\\
-b^\ast & a^\ast
\end{bmatrix},
\qquad
|a|^2+|b|^2=1 .
\label{eq:supp-s36}
\end{equation}
The remaining sign ambiguity \(J\rightarrow -J\) leaves the Mueller matrix unchanged.
The rotation angle \(\vartheta\) is determined from
\begin{equation}
    \vartheta
    =
    \arccos\left[
    \frac{\operatorname{Tr}(M)-1}{2}
    \right].
\end{equation}
For \(0<\vartheta<\pi\), the corresponding rotation axis is
\begin{equation}
    \vec u
    =
    \frac{1}{2\sin\vartheta}
    \begin{bmatrix}
        M_{zy}-M_{yz}\\
        M_{xz}-M_{zx}\\
        M_{yx}-M_{xy}
    \end{bmatrix}.
\end{equation}
In the form of Eq.~\eqref{eq:supp-s36}, we choose
\begin{equation}
\begin{aligned}
 a
 &=
 \cos\frac{\vartheta}{2}
 -\mathrm{i}u_z\sin\frac{\vartheta}{2},\\
 b
 &=
 -\left(u_y+\mathrm{i}u_x\right)
 \sin\frac{\vartheta}{2}.
\end{aligned}
\end{equation}
Equivalently, in terms of the matrix elements of \(M\),
\begin{equation}
    \begin{aligned}
        a
        &=
        \cos\frac{\vartheta}{2}
        -
        \mathrm{i}
        \frac{M_{yx}-M_{xy}}{2\sin\vartheta}
        \sin\frac{\vartheta}{2},\\
        b
        &=
        \left(
        -\frac{M_{xz}-M_{zx}}{2\sin\vartheta}
        -
        \mathrm{i}
        \frac{M_{zy}-M_{yz}}{2\sin\vartheta}
        \right)
        \sin\frac{\vartheta}{2}.
    \end{aligned}
\end{equation}
With this convention,
\begin{equation}
    |a|^2+|b|^2=1.
\end{equation}

The cases \(\vartheta=0\) and \(\vartheta=\pi\) are singular for the above axis extraction formula. 
For \(\vartheta=0\), one may choose \(J=I\). 
For \(\vartheta=\pi\), the rotation axis should be obtained separately, for example, as the eigenvector of \(M\) with eigenvalue \(1\).

\subsubsection{Jones to Mueller}

Conversely, if the Jones matrix is written in the form of Eq.~\eqref{eq:supp-s36}, with \(|a|^2+|b|^2=1\), its corresponding three-component Mueller matrix is
\begin{equation}
    M =
    \begin{bmatrix}
        1 - 2 [\Re(b)^2 + \Im(a)^2]
        &
        2[\Im(b)\Re(b) + \Re(a)\Im(a)]
        &
        2[\Im(b)\Im(a) - \Re(a)\Re(b)]
        \\
        2[\Im(b)\Re(b) - \Re(a)\Im(a)]
        &
        1 - 2 [\Im(b)^2 + \Im(a)^2]
        &
        2[\Re(b)\Im(a) + \Re(a)\Im(b)]
        \\
        2[\Im(b)\Im(a) + \Re(a)\Re(b)]
        &
        2[\Re(b)\Im(a) - \Re(a)\Im(b)]
        &
        1 - 2 [\Im(b)^2 + \Re(b)^2]
    \end{bmatrix}.
\end{equation}

\subsection{Axis-angle coordinates}

Each measured \(SO(3)\) Mueller matrix is first converted to the Jones representative defined in Eq.~\eqref{eq:supp-s36}. 
We then introduce a reduced axis-angle parametrization of this representative,
\begin{align}
\tilde{\epsilon}
&= 2\arccos \Re(a), \nonumber\\
\tilde{\alpha}
&= -\frac{\Re(b)}{\sin(\tilde{\epsilon}/2)}, \nonumber\\
\tilde{\beta}
&= \frac{\Im(a)}{\sin(\tilde{\epsilon}/2)}, \nonumber\\
\tilde{\gamma}
&= -\frac{\Im(b)}{\sin(\tilde{\epsilon}/2)} ,
\label{eq:supp-reduced-jones-parameters}
\end{align}
for \(\sin(\tilde{\epsilon}/2)\neq0\). \(\tilde{\epsilon}\) is a reduced rotation angle, and \((\tilde{\alpha},\tilde{\beta},\tilde{\gamma})\) define a unit vector. 
The tildes emphasize that these quantities are not physical birefringence parameters, but have similar physical implications \cite{Chartier01,Tentori13}.

We write
\begin{equation}
\tilde{\mathbf u}
=
[\tilde{\alpha},\tilde{\beta},\tilde{\gamma}]^{\mathsf T},
\qquad
\|\tilde{\mathbf u}\|^2
=
\tilde{\alpha}^2+\tilde{\beta}^2+\tilde{\gamma}^2
=
1 .
\end{equation}

The parameters \((\tilde{\epsilon},\tilde{\alpha},\tilde{\beta},\tilde{\gamma})\) are not unique. 
They depend on the chosen Jones representative and on the sign convention used in calculating the axis-angle parameters in Eq.~\eqref{eq:supp-reduced-jones-parameters}. 
Before interpolation, we therefore have to choose a locally continuous branch. 
We simultaneously replace
\[
\tilde{\epsilon}\rightarrow -\tilde{\epsilon},
\qquad
\tilde{\mathbf u}\rightarrow -\tilde{\mathbf u}. 
\]
This operation is a parameter branch correction, not a change from \(J\) to \(-J\), because the product \(\tilde{\mathbf u}\sin(\tilde{\epsilon}/2)\) is invariant.  

For consecutive wavelength points, we choose the sign of the current axis by local continuity. 
If
$\left|\tilde{\mathbf u}_{j}-\tilde{\mathbf u}_{j-1}\right|
>
\left|-\tilde{\mathbf u}_{j}-\tilde{\mathbf u}_{j-1}\right|$,
equivalently if
$\tilde{\mathbf u}_{j-1}\cdot \tilde{\mathbf u}_{j}<0$, we replace
\[
\tilde{\mathbf u}_{j}\rightarrow -\tilde{\mathbf u}_{j},
\qquad
\tilde{\epsilon}_{j}\rightarrow -\tilde{\epsilon}_{j}.
\]
After this axis sign continuity step, \(\tilde{\epsilon}\) is unwrapped as a \(2\pi\)-periodic angular coordinate, so that neighboring values lie on a continuous branch. 
A change \(\tilde{\epsilon}\rightarrow\tilde{\epsilon}+2\pi\) changes the Jones representative from \(J\) to \(-J\), but both representatives correspond to the same Mueller matrix.

\subsection{Auxiliary-wavelength interpolation}

Guided by the inverse-wavelength scaling of simple retarder phase delays, we use \(1/\lambda\) as the interpolation coordinate. 
Suppose that the parameters are known at two auxiliary wavelengths
\(\lambda_1\) and \(\lambda_2\), and that the desired signal wavelength is \(\lambda_0\). 
We use inverse wavelength as the interpolation coordinate and define
\begin{equation}
x_j=\frac{1}{\lambda_j},
\qquad j\in\{0,1,2\}.
\end{equation}
The interpolation parameter is therefore
\begin{equation}
t
=
\frac{x_0-x_1}{x_2-x_1}
=
\frac{\lambda_0^{-1}-\lambda_1^{-1}}
{\lambda_2^{-1}-\lambda_1^{-1}} .
\end{equation}
For \(\lambda_0\) between \(\lambda_1\) and \(\lambda_2\), this gives \(0\leq t\leq 1\), provided that the wavelengths are ordered in the chosen interpolation coordinate.

The reduced rotation angle is then interpolated linearly in \(1/\lambda\):
\begin{equation}
\tilde{\epsilon}(\lambda_0)
=
(1-t)\tilde{\epsilon}(\lambda_1)
+
t\tilde{\epsilon}(\lambda_2).
\end{equation}

The unit axis \(\tilde{\mathbf u}\) is interpolated using spherical linear interpolation (SLERP) \cite{Shoemake85}. 
Let
\begin{equation}
\mathbf u_1=\tilde{\mathbf u}(\lambda_1),
\qquad
\mathbf u_2=\tilde{\mathbf u}(\lambda_2),
\end{equation}
with
$|\mathbf u_1|=|\mathbf u_2|=1$ .
The angle between the two unit vectors is
\begin{equation}
\Omega
=
\arccos\left(\mathbf u_1\cdot\mathbf u_2\right).
\end{equation}
For \(0<\Omega<\pi\), the interpolated axis is
\begin{equation}
\tilde{\mathbf u}(\lambda_0)
=
\frac{\sin\left[(1-t)\Omega\right]}{\sin\Omega}\,
\mathbf u_1
+
\frac{\sin\left(t\Omega\right)}{\sin\Omega}\,
\mathbf u_2 .
\end{equation}
This construction gives
$\tilde{\mathbf u}(\lambda_1)=\mathbf u_1, \tilde{\mathbf u}(\lambda_2)=\mathbf u_2$,
and preserves
$|\tilde{\mathbf u}(\lambda_0)|=1$ .

The case \(\Omega=\pi\) is not uniquely defined, because infinitely many great circles connect two opposite points on the unit sphere. 
In the present measurements, this case does not occur; if it did, additional information would be required to select the interpolation plane. 

\section{Wavelength dependence of polarization transformations}

In this section, we use the axis-angle representation introduced above, together with directly measured Mueller matrices, to characterize the wavelength dependence of several optical link components. 
We first examine the error caused by determining the polarization compensation at one wavelength and applying it at another. 
We then test wavelength interpolation for moderately wavelength-dependent devices under test (DUTs), and finally consider a strongly wavelength-dependent fiber spool DUT. 

\subsection{Single-wavelength calibration transfer and spectral mismatch}

To motivate the use of two auxiliary wavelengths, we first tested whether a Mueller-matrix calibration measured at one wavelength could be transferred directly to another wavelength. 
We reconstructed the link-segment Mueller matrices at \(\SI{635}{nm}\) and used the resulting compensating matrix at signal
wavelengths of either \(\SI{635}{nm}\) or \(\SI{670}{nm}\). 
The test was performed for three DUTs: a \(\SI{1}{m}\) 630HP fiber segment, a \(\SI{5}{m}\) 630HP fiber mounted on a Thorlabs MPC320 motorized fiber polarization controller (FPC), and a Yogel \(1\times8\) optical switch module with 630HP fiber connections. 
The residual polarization mismatch was quantified using the polarization-induced quantum bit error rate described in the main text: 
\begin{equation}
    {\rm QBER}_{\rm pol}
    =
    \frac{
        1-\Vec{S}\cdot\Vec{S}'
    }{2}, 
    \label{eq:supp-polarization-qber}
\end{equation}
where $\vec S$ is the expected Stokes vector and $\vec S'$ is the actually measured. 

The results are summarized in Table~\ref{tab:supp-table2}. 
\begin{table}[t]
\caption{
\label{tab:supp-table2}
Polarization-induced excess QBER after compensation using Mueller matrices measured at the fixed calibration wavelength \(\SI{635}{nm}\). 
The signal wavelength \(\lambda_{\rm sig}\) was either \(\SI{635}{nm}\) or \(\SI{670}{nm}\). 
Input polarization azimuths were scanned from \(\SI{0}{\degree}\) to \(\SI{179}{\degree}\) in \(\SI{1}{\degree}\) steps.
}
\begin{ruledtabular}
\begin{tabular}{cccc}
\(\lambda_{\rm sig}\) (nm) &
\multicolumn{3}{c}{Excess QBER (\%)}\\
\cline{2-4}
& 1 m fiber & Fiber paddles & Optical switch\\
\colrule
635 & $0.055\pm0.043$ & $0.18\pm0.07$ & $0.13\pm0.09$\\
670 & $0.24\pm0.11$ & $17.97\pm0.96$ & $4.70\pm0.53$\\
\end{tabular}
\end{ruledtabular}
\end{table}
At the calibration wavelength, all tested DUTs are compensated to low excess QBER. 
After transfer from \(\SI{635}{nm}\) to \(\SI{670}{nm}\), however, the residual mismatch becomes strongly device dependent. 
It remains small for the short fiber but increases for the fiber on paddles and the optical switch. 
This shows that Mueller matrices cannot generally be treated as wavelength independent. 

A qualitative picture is that a real fiber link acts as a concatenation of many weak birefringent sections, each with its own local axes and retardance. 
Changing the wavelength modifies the retardance of these sections and the effective birefringence parameters. 
The product of the wavelength dependent local transformations therefore gives a wavelength dependent Mueller matrix for the full link. 
The extent of this spectral variation is device dependent and can be larger for longer or more strongly perturbed fiber paths, such as the fiber on paddles, than for a short fiber segment. 

\begin{figure}[t]
    \centering
    \includegraphics[width=0.5\linewidth]{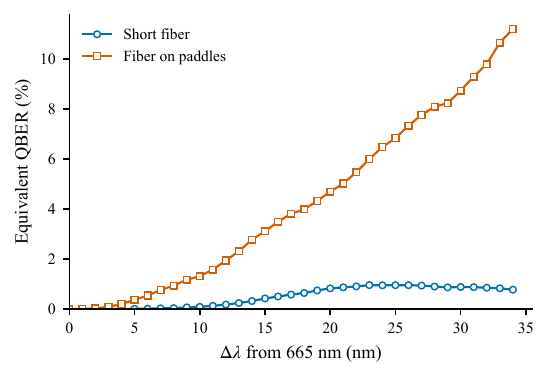}
    \caption{
    Equivalent QBER calculated from Eq.~\eqref{eq:supp-mueller-difference-qber} for Mueller matrices measured at different wavelengths, using \(\SI{665}{nm}\) as the reference wavelength. 
    The ``short fiber'' DUT is a \(\SI{1}{m}\) 630HP fiber segment; the ``fiber on paddles'' DUT is a \(\SI{5}{m}\) 630HP fiber mounted on a FPC.
    }
    \label{fig:supp-fiber_qber_vs_wv}
\end{figure}

To visualize this wavelength dependence, we compare Mueller matrices measured at different wavelengths. 
Taking \(\lambda_0=\SI{665}{nm}\) as the reference wavelength, we define a QBER-like spectral polarization mismatch metric with the same Stokes overlap form as Eq.~\eqref{eq:supp-polarization-qber},
\begin{equation}
{\rm QBER}_{\rm spec}(\lambda;\lambda_0)
=
\frac{1}{2}
\sum_{j\in\{H,D\}}
\frac{
1-\vec m_j(\lambda_0)\cdot \vec m_j(\lambda)
}{2},
\label{eq:supp-mueller-difference-qber}
\end{equation}
where \(\vec m_j(\lambda)\) denotes the measured output Stokes vector, equivalently the corresponding Mueller matrix column, for input state \(j\). 
This quantity quantifies the spectral change in the measured Mueller matrix.
It is the average of Eq.~\eqref{eq:supp-polarization-qber} over the two input states \(H\) and \(D\), with the Stokes vector measured at \(\lambda_0\) taken as the reference. 
Figure~\ref{fig:supp-fiber_qber_vs_wv} shows that the fiber on paddles exhibits a much larger increase of \({\rm QBER}_{\rm spec}(\lambda;\lambda_0)\) with wavelength separation than the short fiber. 
This shows that the spectral variation of the Mueller matrix is device dependent and motivates the auxiliary wavelength interpolation method introduced below.

\subsection{Moderately wavelength dependent DUTs}

To characterize the wavelength dependence of the polarization transformation, we measured the Mueller matrices of 30 DUT configurations at multiple wavelengths. 
The main wavelength scan covered \SI{665}{nm} to \SI{699}{nm} in \SI{1}{nm} steps using a Coherent Chameleon Discovery MX laser. 
An additional measurement at \SI{635}{nm} was performed using a QL63F5SA diode laser. 
The DUT was a \SI{5}{m} length of 630HP single-mode fiber installed on a Thorlabs MPC320 motorized fiber polarization controller. 
Different paddle positions were used to generate different polarization transformations.

For each DUT configuration and wavelength, the measured Mueller matrix was converted to the axis-angle representation \((\tilde{\epsilon},\tilde{\alpha},\tilde{\beta},\tilde{\gamma})\) defined above. 
We then applied the wavelength interpolation procedure. 
The interpolation was constructed from the two endpoint measurements at \SI{665}{nm} and \SI{699}{nm}. 
The same two-point model was also extrapolated toward \SI{635}{nm} to test the possibility of using auxiliary wavelengths outside the interval of interest.

\begin{figure}
    \centering
    \includegraphics[width=0.9\textwidth]{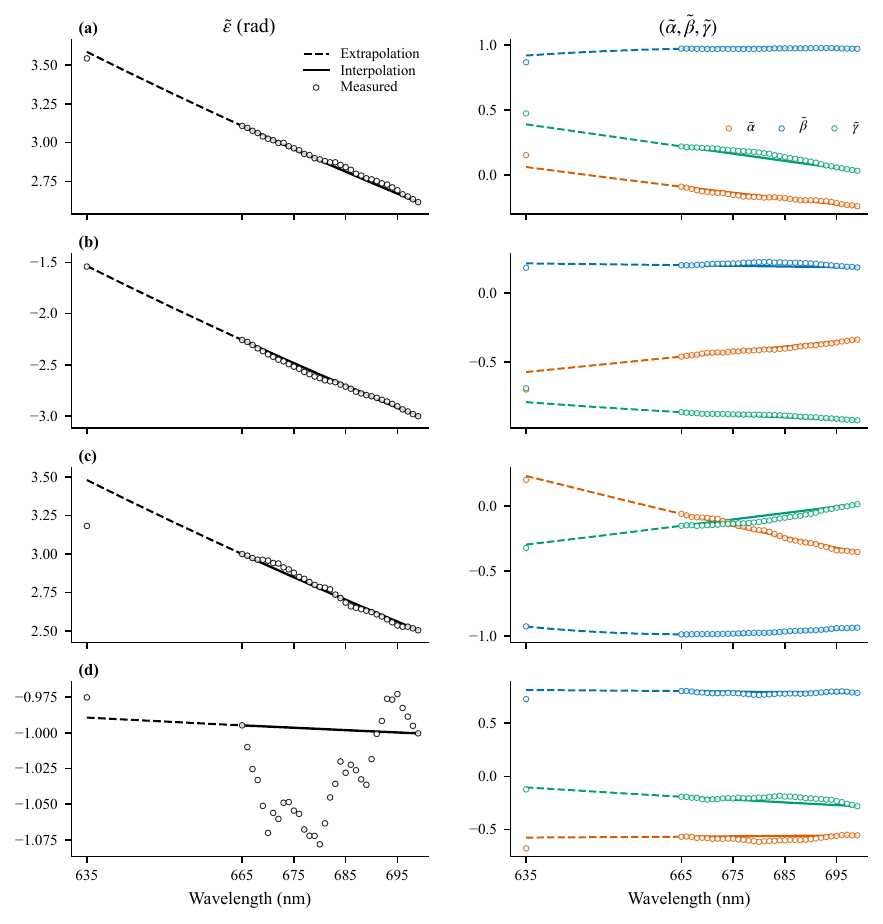}
    \caption{
    Representative wavelength dependence of the axis-angle parameters for four moderately wavelength dependent DUT configurations. 
    In each row, the left panel shows the reduced angle \(\tilde{\epsilon}\), and the right panel shows the unit vector components \((\tilde{\alpha},\tilde{\beta},\tilde{\gamma})\). 
    Open circles denote measured data. 
    Solid curves show the axis-angle parameter interpolation obtained from the endpoint measurements at \SI{665}{nm} and \SI{699}{nm}. 
    Dashed curves show the corresponding extrapolation toward \SI{635}{nm}. 
    \textbf{(a)} and \textbf{(b)} Examples where the interpolation within \SI{665}{nm} to \SI{699}{nm} follows the measured parameters well, while the extrapolated unit vector components show larger deviations at \SI{635}{nm}.
    \textbf{(c)} Example where the unit vector components are extrapolated comparatively well, but the extrapolation of \(\tilde{\epsilon}\) shows a larger deviation.
    \textbf{(d)} Example with stronger curvature and non-monotonic behavior, leading to a larger interpolation error within the measured wavelength interval.
    }
    \label{fig:supp_jones_representative}
\end{figure}

Representative dispersion curves for four DUT configurations are shown in Fig.~\ref{fig:supp_jones_representative}. 
For many DUT configurations, the axis-angle parameters vary smoothly over the measured wavelength interval, and the two-point interpolation reproduces the measured data at intermediate wavelengths well. 
However, some configurations show stronger curvature or non-monotonic behavior. 
In such cases, the two-point interpolation leaves a larger residual error. 
These observations support the use of auxiliary wavelength interpolation for moderately dispersive polarization transformations, while also showing that its accuracy depends on the spectral smoothness of the DUT Mueller matrix.

To quantify how this wavelength dependence affects compensation based on auxiliary wavelengths, we evaluated the residual error that would remain if the signal wavelength segment matrices were estimated from auxiliary wavelength data. 

In the cases where two auxiliary wavelengths are used, the signal wavelength segment matrices \(\hat M_{\alpha}\) and \(\hat M_{\beta}\) are estimated with interpolation. 
In the cases with only one auxiliary wavelength, no interpolation is performed; the Mueller matrices measured at the listed auxiliary wavelength are directly transferred to the signal wavelength. 
The corresponding estimated compensator is
\begin{equation}
    \hat M
    =
    \hat M_{\beta}^{\mathsf T}
    \hat M_{\alpha}^{\mathsf T}.
\end{equation}

We combine this estimated compensator with the independently measured signal wavelength segment matrices,
\begin{equation}
    M_{\rm tot}
    =
    M_{\beta,s}
    \hat M
    M_{\alpha,s}.
\end{equation}
The residual mismatch of the \(H\) and \(D\) basis states defines the QBER induced by interpolation 
\begin{equation}
{\rm QBER}_{\rm int}
=
\frac{1}{2}
\sum_{j\in\{H,D\}}
\frac{
1-\vec S_j\cdot M_{\rm tot}\vec S_j
}{2}.
\end{equation}

Table~\ref{tab:supp-table3} reports the average, standard deviation, maximum, and minimum of \({\rm QBER}_{\rm int}\) over the 30 DUT configurations.
\begin{table*}
\caption{\label{tab:supp-table3}
Excess QBER estimated from measured wavelength dependent Mueller matrices. 
The statistics are calculated over 30 paddle positions of the fiber
polarization controller. 
``Aux. 1'' and ``Aux. 2'' denote the wavelengths used to estimate the signal wavelength Mueller matrices. 
In the rows with single wavelength, only one auxiliary wavelength is used and the Mueller matrix measured at that wavelength is assumed to apply at the signal wavelength. 
The reported average, standard deviation, maximum, and minimum values are given in percent. 
}
\begin{ruledtabular}
\begin{tabular}{c ccc dddd}
 & \multicolumn{3}{c}{Wavelength (nm)}
 & \multicolumn{4}{c}{\({\rm QBER}_{\rm int}\) (\%)}\\
\cline{2-4}\cline{5-8}
 & Aux. 1 & Aux. 2 & Signal
 & \multicolumn{1}{c}{Avg.}
 & \multicolumn{1}{c}{Std.}
 & \multicolumn{1}{c}{Max.}
 & \multicolumn{1}{c}{Min.}\\
\hline
\multirow{3}{*}{\shortstack{Single auxiliary\\ wavelength}} &  670&  --&  635&  10.95&  7.16&  22.83& 0.19\\
         &  665&  --&  680&  1.69&  1.15&  3.52& 0.11\\
         &  695&  --&  680&  1.76&  1.14&  4.83& 0.18\\ \hline
         \multirow{2}{*}{\SI{10}{nm} separation} &  689&  699&  665&  0.42&  0.19&  0.80& 0.10\\
         &  689&  699&  635&  1.26&  0.80&  3.11& 0.06\\ \hline
         \multirow{2}{*}{\SI{30}{nm} separation}&  665&  695&  680&  0.06&  0.03&  0.18& 0.01\\
         &  665&  695&  635&  1.37&  0.73&  2.81& 0.12\\ \hline
        \SI{34}{nm} separation& 665& 699& 682& 0.06& 0.03& 0.15&0.01\\ \hline
         \SI{60}{nm} separation &  635&  695&  665&  0.41&  0.26&  0.96& 0.03\\ 
\end{tabular}
\end{ruledtabular}
\end{table*}
For moderate polarization distortions produced by the fiber polarization controller, interpolation within the measured wavelength interval keeps the estimated excess QBER comparable to the residual errors caused by the present hardware. 
Extrapolation can work for some paddle positions, but it is less reliable and can produce larger errors. 
These results show that the auxiliary wavelength method is most robust when the signal wavelength lies between, and sufficiently close to, the two auxiliary wavelengths.

\subsection{A strongly wavelength dependent DUT}

We also examined a more strongly wavelength dependent DUT consisting of a \SI{100}{m} single-mode fiber wound on a spool. Compared with the shorter fiber DUTs discussed above, this configuration can accumulate a much larger bend-induced retardance. 
As a result, a small change in wavelength can correspond to a substantial change in the effective polarization transformation on the Poincaré sphere.

For this DUT, we did not perform a dense wavelength scan. 
Instead, we measured the Mueller matrices at three wavelengths for two representative auxiliary wavelength configurations. 
In the wide-spacing case, the three wavelengths were provided by separate diode lasers with measured spectral peaks at \SI{658.3}{nm}, \SI{672.3}{nm}, and \SI{686.5}{nm}. 
In the close-spacing case, the wavelengths were obtained by current tuning of the nominal \SI{670}{nm} diode, giving wavelengths near \SI{669}{nm}, \SI{670}{nm}, and \SI{671}{nm}. 

\begin{figure}
    \centering
    \includegraphics[width=0.8\textwidth]{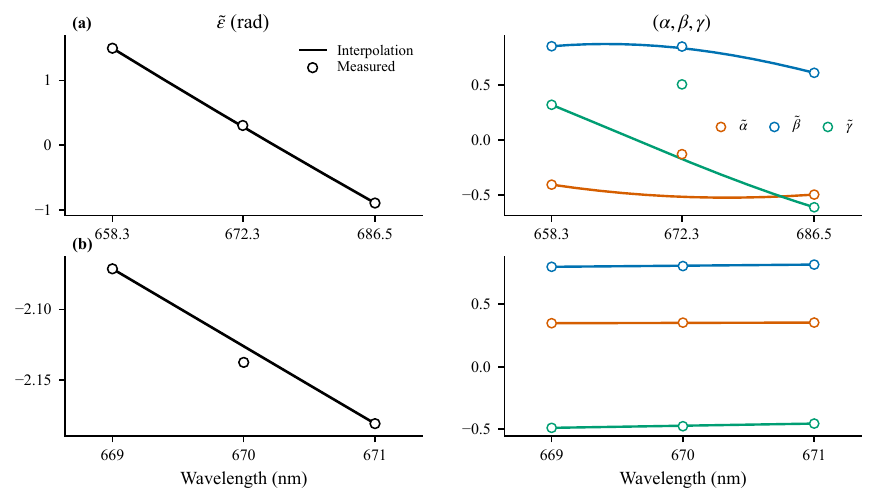}
    \caption{
    Axis-angle parameter interpolation for a strongly wavelength dependent DUT consisting of a \SI{100}{m} single-mode fiber wound on a spool. 
    The two rows correspond to measurements taken under different experimental conditions and should not be interpreted as a consistent wavelength scan of the same unchanged Mueller matrix.
    \textbf{(a)} Wide-spacing case using three diode lasers with measured peak wavelengths of \SI{658.3}{nm}, \SI{672.3}{nm}, and \SI{686.5}{nm}. 
    \textbf{(b)} Close-spacing case obtained by current tuning of the nominal \SI{670}{nm} diode, with wavelengths near \SI{669}{nm}, \SI{670}{nm}, and \SI{671}{nm}. 
    }
    \label{fig:supp_extreme_dut_jones}
\end{figure}

The axis-angle parameters are shown in Fig.~\ref{fig:supp_extreme_dut_jones}. 
In the wide-spacing case, the middle wavelength point deviates from the interpolation results between the two outer wavelengths, indicating that the polarization transformation varies appreciably over this spectral interval. 
In contrast, when the auxiliary wavelengths are separated by only about \SI{2}{nm}, the three measured points are much more nearly collinear in the axis-angle parameter representation. 

\section{Temperature variations and polarization drift}

To characterize temperature-induced polarization drift, we measured the Mueller matrix of the fiber spool DUT placed in a temperature-controlled enclosure. 

\subsection{Daily soil temperature variations}

\begin{figure}[t]
    \centering
    \includegraphics[width=\linewidth]
    {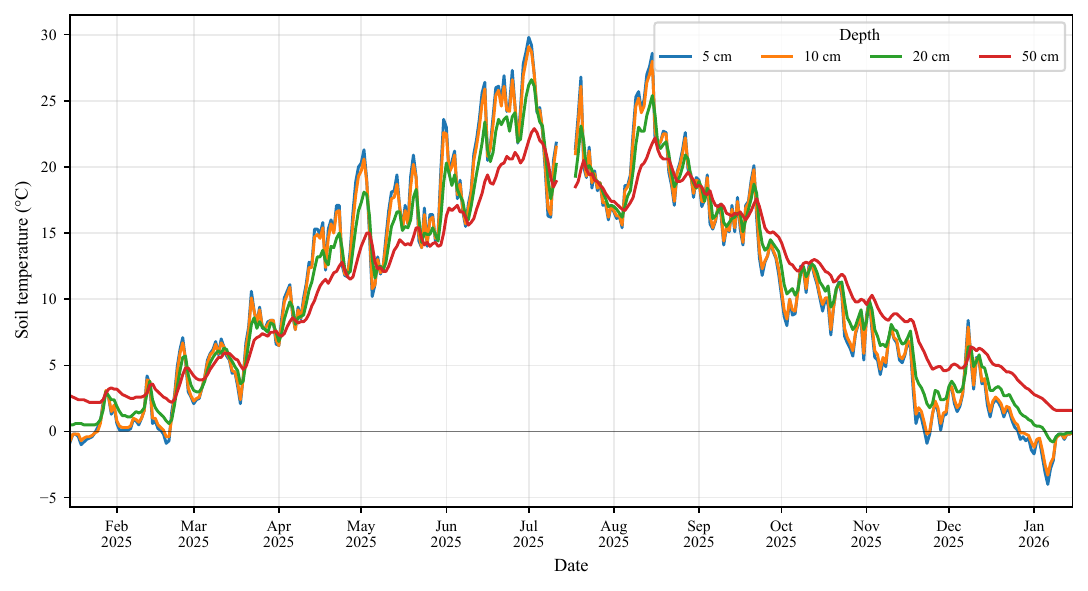}
    \caption{Daily soil temperature at different depths measured at the DWD climate station M\"unchen-Flughafen (station ID 01262) over a one-year interval beginning on January 15, 2025.}
    \label{fig:supp-soil_temperature}
\end{figure}

To determine the temperature range for the experiment, we used data from the Climate Data Center (CDC) of the Deutscher Wetterdienst (DWD)~\cite{DWD26}. 
The climate station is located at Munich Airport, close to the Garching research campus. 
As shown in Fig.~\ref{fig:supp-soil_temperature}, the annual temperature range narrows with increasing depth, as do the temperature variations between consecutive days. 
The temperature at all measured depths remains below \SI{30}{\degreeCelsius}. 
Accordingly, the upper bound of the experimental temperature ramp was set to \SI{30}{\degreeCelsius}. 
The lower bound was limited by the temperature-controlled enclosure, whose cooling function is implemented using a Peltier module. 
Depending on the ambient temperature, we observed that cooling stopped when the temperature reached between \SI{12}{\degreeCelsius} and \SI{16}{\degreeCelsius}. 
We therefore chose a temperature ramp between \SI{15}{\degreeCelsius} and \SI{30}{\degreeCelsius} for the following
experiment. 

\subsection{Temperature-induced Mueller-matrix drift}

A passive drift characterization was performed in a separate measurement from the active auxiliary wavelength feedback experiment in the main text, and was used to quantify the same type of temperature-induced Mueller matrix variation. 
In this measurement, no polarization compensation was applied during the temperature ramp. 
The goal was therefore to measure the passive change of the Mueller matrix with temperature.

We quantify this drift using an equivalent QBER calculated relative to the Mueller matrix measured at \SI{15}{\degreeCelsius}. 
This corresponds to the situation in which a compensation setting is chosen at the beginning of the measurement at \SI{15}{\degreeCelsius}, and then kept fixed while the device temperature changes. 
The temperature dependence is also visualized through the corresponding axis-angle parameters.

\begin{figure}[t]
    \centering
    \includegraphics[width=\linewidth]{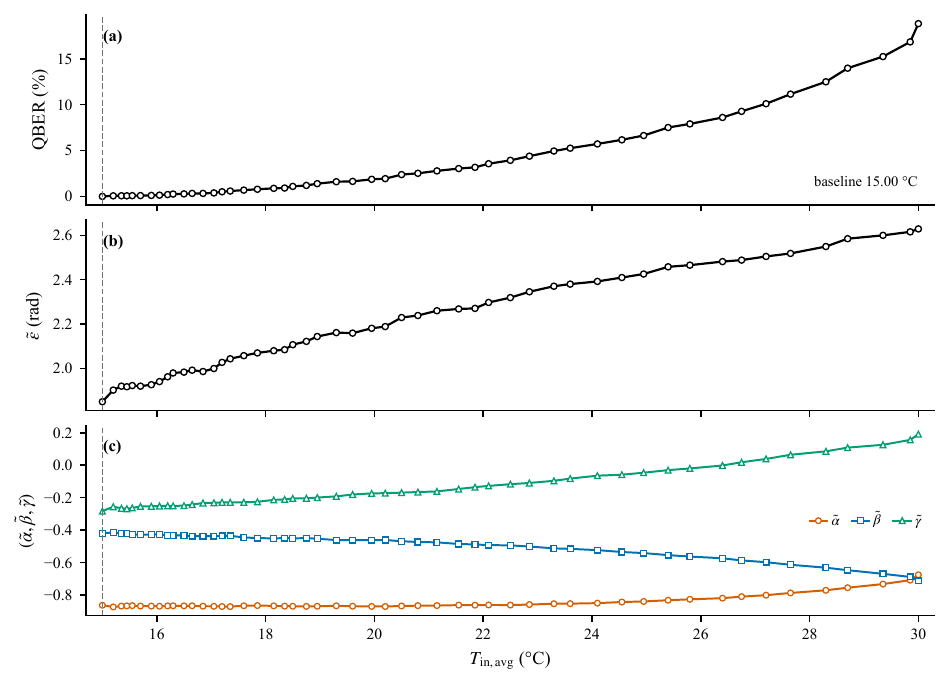}
    \caption{
    Temperature-dependent polarization drift of the \SI{100}{m} fiber spool DUT measured without active compensation during the ramp. 
    The horizontal axis is the average internal enclosure temperature \(T_{\mathrm{in,avg}}\).
    \textbf{(a)} Equivalent QBER calculated from the measured Mueller matrices relative to the baseline Mueller matrix at \SI{15}{\degreeCelsius}. 
    \textbf{(b)} Reduced rotation angle \(\tilde{\epsilon}\).
    \textbf{(c)} Rotation axis components \((\tilde{\alpha},\tilde{\beta},\tilde{\gamma})\). 
    }
    \label{fig:supp-temperature-ramp-qber-jones}
\end{figure}

The results are shown in Fig.~\ref{fig:supp-temperature-ramp-qber-jones}. 
As the temperature changes, the measured Mueller matrix drifts away from its initial value. 
In the absence of active compensation, this drift would lead to a large residual polarization mismatch relative to the initial compensation condition. 
For the data shown here, the equivalent QBER increases from zero at the baseline by construction to values above \SI{15}{\percent} near the end of the ramp.
This passive drift measurement provides a reference for the active compensation experiment in the main text, where auxiliary wavelength feedback maintained the measured residual QBER below \SI{1}{\percent} during temperature variation.

\section{Local feedback procedure used in the long-term experiment}

After the initial auxiliary-wavelength compensation, the link was monitored without repeating the full eight-Stokes segment reconstruction protocol. 
This can prevent disturbing the compensated signal wavelength link. 

For the current wave plate setting
\(\boldsymbol{\theta}=(\theta_+,\theta_1,\theta_2,\theta_3)\), we denote the
total compensated Mueller matrix by
\begin{equation}
    M_{\rm tot}(\lambda;\boldsymbol{\theta})
    =
    M_\beta(\lambda)
    M_{\rm WP}(\lambda;\boldsymbol{\theta})
    M_\alpha(\lambda).
\end{equation}
The target condition is \(M_{\rm tot}(\lambda_s;\boldsymbol{\theta})=I\). 
The feedback metric uses the reduced rotation angle $\tilde{\epsilon}$ extracted from \(M_{\rm tot}\). 
At each iteration, this angle \(\hat{\tilde{\epsilon}}_{\rm tot}(\lambda_s;\boldsymbol{\theta})\) was obtained by measuring at the two auxiliary wavelengths and interpolating to the signal wavelength. 
The feedback metric was
\begin{equation}
    F(\boldsymbol{\theta})
    =
    \left|
    \hat{\tilde{\epsilon}}_{\rm tot}(\lambda_s;\boldsymbol{\theta})
    \right|.
\end{equation}

In each iteration, the Q-Q-Q-H angle-solving algorithm was reapplied to the current Mueller $M_{\rm WP}(\lambda;\boldsymbol{\theta})$ of the wave plates. 
The resulting recalculated value of the additional QWP angle \(\theta_+\) was used as an update target, but the plate was moved only by a small step toward this recalculated value in order to avoid a large change of the compensated link. 
The remaining three angles, \(\theta_1\), \(\theta_2\), and \(\theta_3\), were optimized locally by scanning small angular intervals around their current settings and selecting the values that minimized \(F\). 

\section*{Notes}
We acknowledge the use of Microsoft Copilot with the GPT-5 reasoning model to assist in improving Python scripts used for figure generation and in revising selected passages of the manuscript for readability. All scientific interpretations, data analyses, and final writing decisions are made by the authors.

\end{document}